%% file: paper.tex
\documentclass[letterpaper,twocolumn,10pt]{article}

\usepackage{usenix2019_v3}
\usepackage{tabularx,enumitem}
\usepackage{booktabs,amsmath,amsfonts,amssymb}
\usepackage{endnotes,hyperref,color}
\usepackage[noend]{algpseudocode}
\usepackage[sort]{cite}
\usepackage{textcomp,gensymb,xspace,hyphenat}
\usepackage{authblk}

\usepackage[labelformat=simple,skip=1pt]{caption}
\usepackage{amsmath}
\usepackage{subcaption}
\usepackage{amsthm}
\usepackage{graphicx}

\usepackage[ruled,vlined,linesnumbered]{algorithm2e}

\usepackage{xspace,hyphenat}
\usepackage{listings,xcolor}
\usepackage{tikz}
\usepackage{tcolorbox}
\usepackage{pifont}
\usepackage{times}
\usepackage{makecell}

\usepackage[left=2cm,right=2cm,bottom=2.5cm,headsep=10pt]{geometry}  

\microtypecontext{spacing=nonfrench}

\newcolumntype{P}[1]{>{\centering\arraybackslash}p{#1}}
\newcolumntype{M}[1]{>{\centering\arraybackslash}m{#1}}

\newcommand{\ie}{{\itshape i.e.}\xspace}
\newcommand{\eg}{\emph{e.g.}\xspace}

\newcommand{\parahead}[1]{\vspace*{4.5pt plus 2pt minus 2pt}\noindent %
  {\bfseries #1}}

\newcommand{\parabreak}{\vspace*{2ex plus 0.5ex minus 0.5ex}\noindent}
\renewcommand{\paragraph}[1]{\parahead{#1}}

\newcommand{\systemname}{Dashlet}

\begin{document}

\twocolumn[\begin{@twocolumnfalse}

\begin{centering}
{\large \textbf{\systemname: Taming Swipe Uncertainty for Robust Short Video  Streaming \vspace{-0.1cm}}\\\vspace{6pt} \large Zhuqi Li, Yaxiong Xie, Ravi Netravali, Kyle Jamieson \vspace{6pt}}


\end{centering}

\vspace{\baselineskip}

\end{@twocolumnfalse}]

\setitemize{itemsep=1pt,topsep=2pt,parsep=1pt,partopsep=0pt,leftmargin=1.5em}
\setenumerate{itemsep=1pt,topsep=2pt,parsep=1pt,partopsep=0pt,leftmargin=1.5em}
\setdescription{itemsep=1pt,topsep=2pt,parsep=1pt,partopsep=0pt,leftmargin=0pt}
\setlist{itemsep=1pt,parsep=1pt}


\input{Abs}
\input{Intro}

\input{tiktok}
\input{user_swipes}

\input{Design}
\input{Eval}

\input{related}
\input{concl}

\let\oldbibliography\thebibliography
\renewcommand{\thebibliography}[1]{%
  \oldbibliography{#1}%
  \setlength{\itemsep}{0pt}%
}

\cleardoublepage{}
\bibliographystyle{acm}
\begin{raggedright}
\bibliography{reference}
\end{raggedright}


\end{document}

%% file: Abs.tex
\section*{Abstract}

Short video streaming applications have recently gained
substantial traction, but the non-linear video presentation they afford 
swiping users fundamentally changes the problem of maximizing user 
quality of experience in the face of the vagaries of network throughput
and user swipe timing.
This paper describes the design and implementation of \systemname, 
a system tailored for high quality of experience in short video 
streaming applications.  With the insights we glean from
an in-the-wild TikTok performance study and a user study focused on swipe patterns, 
\systemname{} 
proposes a novel out-of-order video chunk pre-buffering 
mechanism that leverages a simple, \emph{non} machine learning-based 
model of users' swipe statistics to determine the pre\hyp{}buffering order and bitrate.   
The net result is a system that achieves 77--99\%
of an oracle system's QoE and outperforms TikTok by 43.9-45.1$\times$, while also reducing by 30\% the number of bytes wasted on downloaded video that is never watched.

%% file: Intro.tex
\section{Introduction}
\label{s:intro}

Short video streaming applications like TikTok and YouTube Shorts have rapidly risen in popularity, attracting billions of active users per month~\cite{tikTokrevenue,tiktokpeople,youtubeshort} and consistently topping popularity lists for mobile apps~\cite{tiktokstats2}. Unlike typical video streaming, the median duration of short videos is around 14 seconds~\cite{chen2019study}. During operation, these apps generate an ordered playlist of short videos (\emph{e.g.}, based on a search or user\hyp{}specific recommendations), 
and users watch them serially, with the ability to swipe from one to the next at any time. To provide an immersive experience and keep users engaged, short video streaming applications should minimize the video rebuffering time and maximize the video bitrate, which is modeled by quality-of-experience (QoE)~\cite{svideoqoe1,svideoqoe2,svideoqoe3,tiktokstats}.

\begin{figure}
    \centering
    \includegraphics[width=\linewidth]{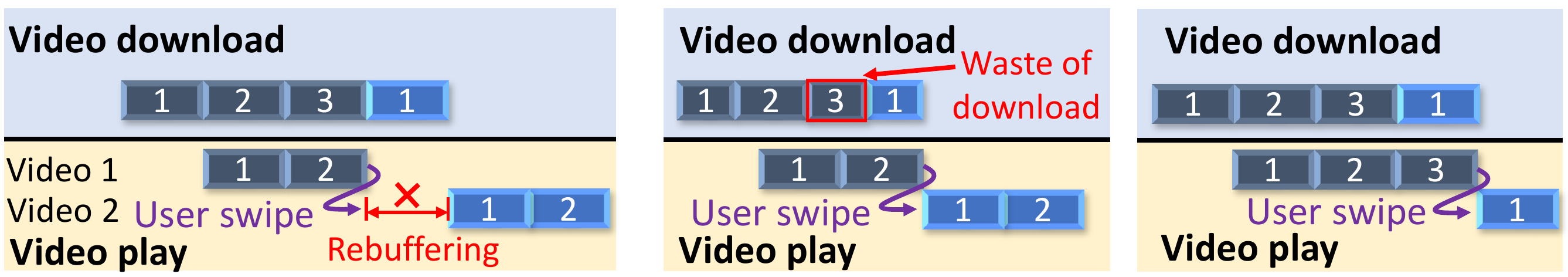}
    \caption{In short video apps, user swipes dictate the playing order of video chunks (and thus, the optimal chunk downloading order).}
    \label{fig:intro_view_seq}
    \vspace{-10pt}
\end{figure}

Although the aforementioned goals are consistent with those in traditional video streaming scenarios, existing ABR 
algorithms~\cite{Pensieve,MPC,oboe,puffer,bufferbased} are ill\hyp{}suited for interactive, short videos. The reason is that predicting user swipes is difficult, and swipe times dictate both which video content will be viewed and when during a session. However, existing algorithms assume that the user will watch
content sequentially to completion, and will hence buffer chunks (i.e., multi-second blocks of video) in that order. The deleterious effects, shown in Figure~\ref{fig:intro_view_seq}, are twofold: 
\textbf{(1)}~many chunks may be downloaded in the current video 
but never viewed if the user swipes before their playback,
wasting resources and adding delays for the chunks that are 
required, and \textbf{(2)}~users may swipe to the next video and incur significant rebuffering because that video's chunks 
have not been downloaded yet.  

The fundamental challenge is that there are far too many possible chunk viewing sequences---the user may swipe at any position in each short video, and expects seamless (\ie, no stalls) playback for both the current video, and the next one upon a swipe.
The problem thus becomes how to find (at any time during playback) a buffering sequence of chunks in this large search space that maximizes QoE by simultaneously minimizing rebuffering time and wasted bandwidth.

To understand how commercial short streaming platforms attempt
to address these challenges, we conducted a detailed 
examination of TikTok in the wild (\S\ref{ss:measurement-tt}). Our key finding is that 
TikTok does download chunks out of order, but follows a 
generic algorithm that hedges against immediate rebuffering 
in the face of fast user swipes (it always pre\hyp{}buffers 
the first chunk for the next five videos regardless of network
conditions, user patterns, and\fshyp{}or video). This, however, entails 
substantial QoE penalties and wasted data consumption, as we will show via results from our own study of user swipe patterns across two distinct sets of users on a 
college campus and Amazon Mechanical Turk (\S\ref{s:measurement-swipe}). Specifically, we find substantial heterogeneity in the swipe patterns across users, with each warranting a different chunk downloading strategy.

A na\"{\i}ve solution would be to simply predict user 
swipes---if accurate, this would reduce the problem to a
traditional streaming setting since chunk viewing sequences would be known a priori. However, predicting 
user behavior in interactive applications has consistently proven to be difficult~\cite{parsec,qian2018flare,pocketweb}. 
Instead, we take a more fundamental approach that is rooted in an understanding of where swipe predictions are actually helpful (and actionable).

We present \textbf{\systemname{}}, a new video streaming 
algorithm for short video applications (\S\ref{s:design}). The underlying 
insight behind \systemname{} is that application playback constraints predetermine the relative priorities between many chunks that are candidates for buffering. More specifically, \textbf{(1)}~later chunks in a video are only reachable via earlier ones, and \textbf{(2)}~later videos are only reachable via swipes from earlier ones. To prioritize among the remaining chunks, \emph{e.g.}, the next chunk in a given video \emph{vs.} the first chunk in the next video, only coarse grained information about swipe timings in videos is required. We show, via our user study, that although users tend to exhibit multimodal swipe patterns (complicating chunk prioritization) across videos, distributions from aggregating users' swipes \emph{per video} provide a clear enough signal about which mode to expect. This information is readily available to current short video platforms, and our finding is spiritually aligned with past studies that highlight similarities in user engagement for certain video content~\cite{sensei,salientvr}.

Building on this, \systemname{} develops functions that 
characterize the expected rebuffering time for each potential 
chunk that could be downloaded, as a continuous function 
over both the expected download and playback times.  These functions embed the aforementioned inter\hyp{}chunk relationships, as well as rough swipe likelihoods at video start and end. 
Using these functions, \systemname{} employs a greedy algorithm
to determine the set of ordered chunks that should be downloaded
in the current time horizon to minimize expected 
rebuffering delays for a given network estimate and 
across potential viewing sequences. This buffer sequence then feeds directly into a traditional ABR 
algorithm, which determine bitrates for those chunks that maximize
overall QoE. \systemname{} further improves upon existing short video systems by not prematurely binding bit rate decisions across entire  short videos, and not letting the network idle at any point in time. 

We have implemented \systemname{} in the DASH framework~\cite{dashif}, and compare with the TikTok mobile app on a wide range of mobile network conditions, 100 short videos, and real user swipe traces (from our user studies). Across these conditions, we find that \systemname{} achieves QoE values within 77.3-98.6\% of an oracle strategy (based on MPC~\cite{MPC}), outperforming TikTok by 43.9-45.1$\times$ (85.3-246.2\% higher bitrate rewards, 102.4-128.4$\times$ lower rebuffering penalties, and 30\% less wasted bandwidth). Further,  \systemname{} is largely tolerant to errors in swipe distributions: QoE degradations are only 10\% with distribution errors of 50\%. We will open source our datasets and implementation post publication.






%% file: tiktok.tex
\section{A TikTok Case Study}
\label{ss:measurement-tt}

We examine how TikTok, a state-of-the-art short video app, operates. We first describe its basic architecture (\S\ref{s:tiktok_primer}), before analyzing its operation and limitations
(\S\ref{s:tiktok_analysis}).

\input{tiktok_primer}
\input{tiktok_analysis}

\parabreak{}To understand the mismatch between TikTok's generic rule 
and the varying user swipe patterns, 
in the next section, we characterize the swipe patterns across 
real users and videos via two user studies.

%% file: tiktok_primer.tex
\subsection{Short Video Streaming Primer}
\label{s:tiktok_primer}\label{ss:primer}

In contrast to traditional streaming apps that divide video 
into chunks of equal time duration, TikTok instead splits each video 
into size\hyp{}based
chunks.  For each supported bitrate, if the video file is smaller than 
1~MB, TikTok treats the entire video as one chunk; else, the first chunk 
is the first 1 MB, and the entire remaining video is grouped into a second 
chunk. As we will highlight below, the rationale for this chunking strategy
is to enable more stable (and consistent) download delays as TikTok
pre\hyp{}buffers the first chunk for subsequent videos (to cope with swipe 
uncertainty); chunking in terms of bytes eliminates variance 
from variable bitrate encoding.

\begin{figure}
    \centering
        \includegraphics[width=0.82\linewidth]{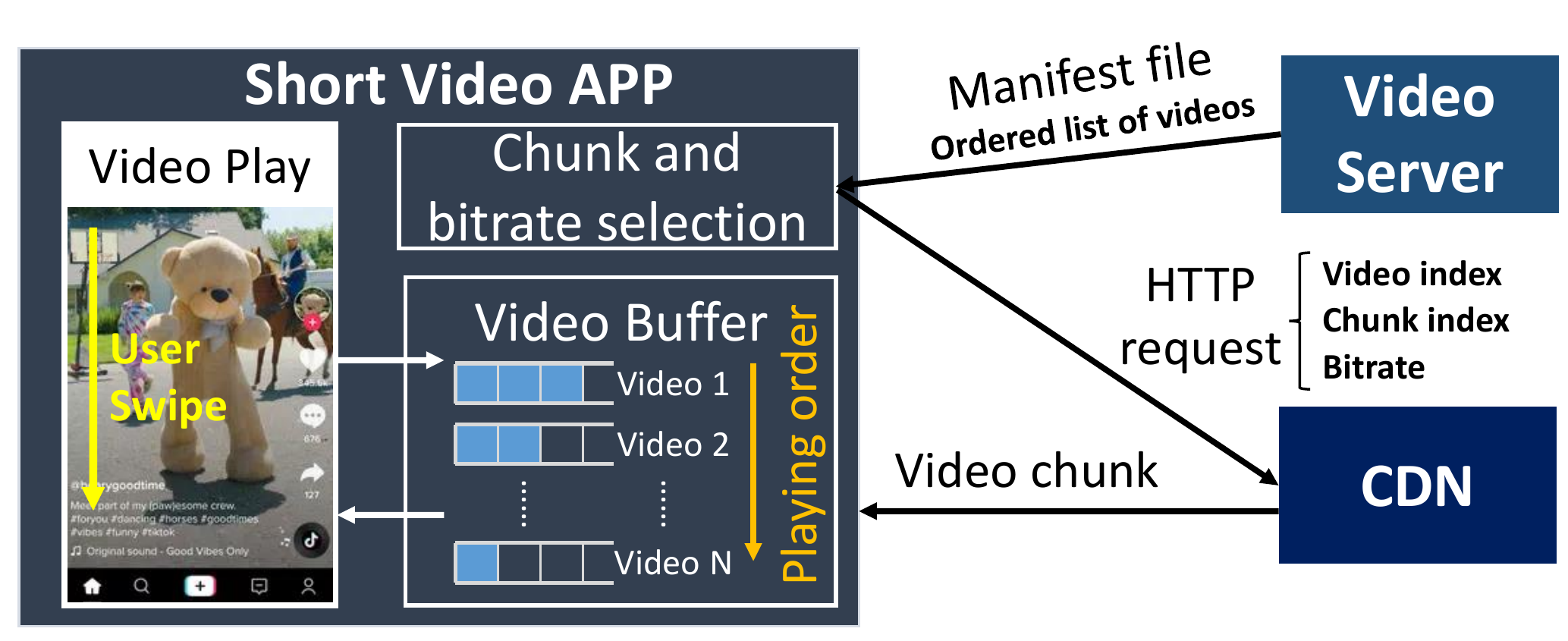}
        \vspace{-3pt}
        \caption{System architecture of TikTok and other short video apps.}
    \label{fig:sys_arch}
    \vspace{-14pt}
\end{figure}

Upon receiving a client session request either via a keyword search or category selection (\emph{e.g.}, recommended videos), the 
server generates an ordered list of short videos to serve (Figure~\ref{fig:sys_arch}). The server then ships a 
\emph{manifest file}
to the client which embeds the URL, as well as information 
about the number of \emph{chunks} 
(multi-second blocks of video) and available bitrates, per video 
in the ordered list. The client operates much like a traditional streaming player (\eg, DASH), maintaining a playback buffer for 
downloaded video and employing an adaptive bitrate 
(ABR) algorithm to determine what chunk to download next, 
when, and at what bitrate.

\begin{figure*}[t]
    \begin{subfigure}[t]{\linewidth}
        \centering
        \includegraphics[width=\linewidth]{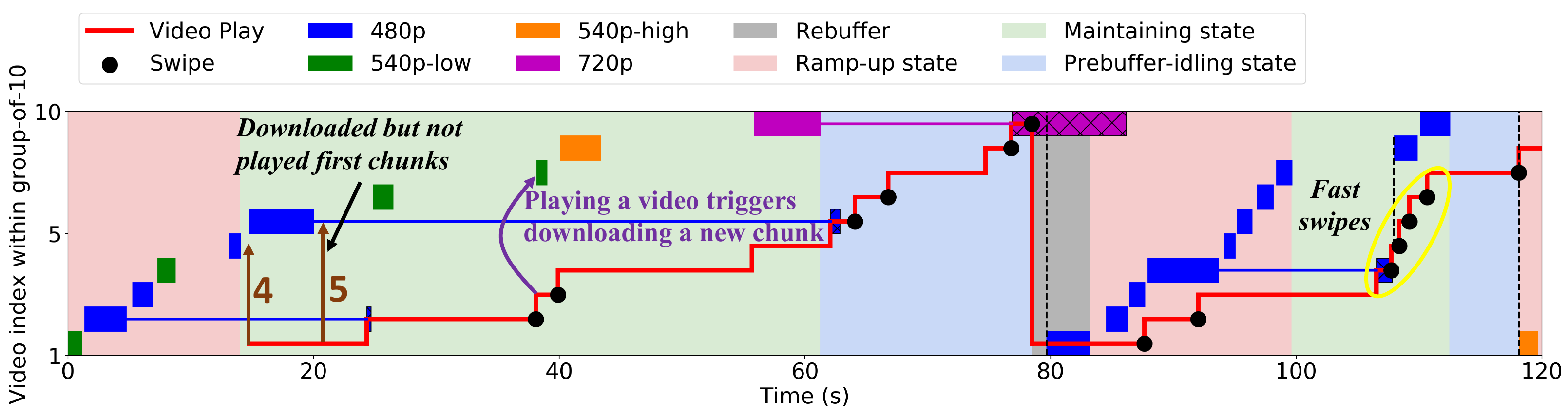}
        \caption{Video chunk downloading and playing timeline: 
        video index within a group-of-10 versus wall clock time. The 
        \textit{left and right edge} of the rectangular boxes 
        respectively represent the downloading start and completion times of a chunk,
        while thin horizontal lines connect first and second chunks (if a second
        chunk exists), and box
        \textit{color} indicates bitrate.  The solid red line
        plots video playback.}
        \label{fig:playback-demo}
    \end{subfigure}
    \begin{subfigure}[t]{\linewidth}
        \centering
        \includegraphics[width=\linewidth]{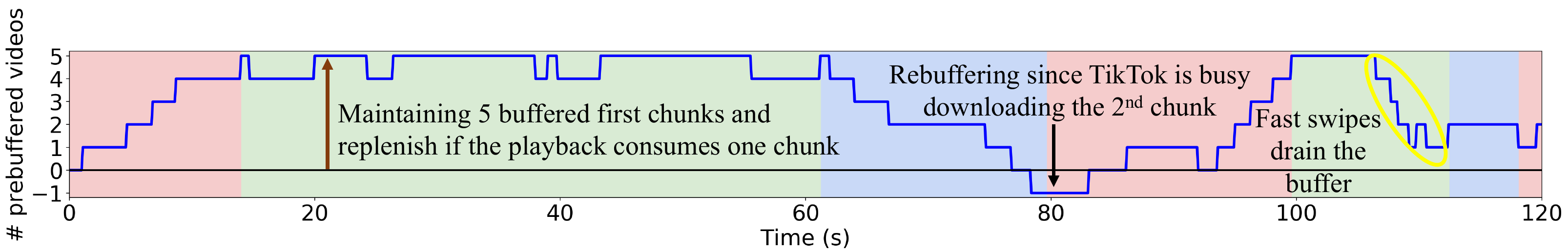}
        \caption{Client\hyp{}side buffer occupancy as a function of time,
        the gap between playback and highest chunk downloaded in \textbf{(a)}.}
        \label{fig:buffer_occupancy}
    \end{subfigure}
    \caption{An illustrative video downloading and video playing trace of TikTok, with
    associated video bitrate and buffer occupancy statistics.}
    \label{fig:playback-demo_all}
    \vspace{-10pt}
\end{figure*}

A key difference between traditional and short video streaming 
is that the client maintains one logical buffer per
video in the server\hyp{}provided manifest file.
Video playback operates sequentially within each logical buffer 
and across buffers (in the specified order); user swipes 
and video completion trigger the playback to move to the 
head of the buffer for the next video. To cope with such 
semantics, ABR algorithms for short videos have the ability 
to download chunks for any of the videos in the manifest file
at any time.

TikTok provides four bitrate options for each video: 
480p, 560p low, 560p high, and 720p, with
bitrate adaptation occurring only at video\hyp{}level (and not chunk\hyp{}level) 
granularity. We hypothesize this is because the first 1~MB of 
a video encoded at different bitrates corresponds to different time durations, 
precluding seamless bitrate switches for the latter chunk, 
\ie, content would be missed or repeated. 
As we will discuss, such constraints significantly limit 
TikTok from adapting to variations in network capacity during user sessions.

To retain flexibility in recommendations, 
each manifest file contains information for an ordered group of only 10 videos. 
The client requests a new manifest file 
after it downloads all the first chunks of the 10 videos 
listed in the current one. 

%% file: tiktok_analysis.tex
\subsection{Analysis of TikTok}
\label{s:tiktok_analysis}


To study TikTok in a controlled and systematic manner, we perform 
our analysis over emulated networks using 
Mahimahi~\cite{netravali2015mahimahi}. We mirror the screen of 
Android TikTok App to a Linux desktop with scrcpy~\cite{scrcpy} and use
the pyautogui tool~\cite{pyAutoGUI} to replay aggregated user
swipe traces that were collected from our user study (described 
in \S\ref{s:measurement-swipe}). During experiments, we use the
mitmproxy~\cite{mitmproxy} to collect and decrypt TikTok's network traffic. 
From the deciphered HTTP messages and headers, we are able to extract 
for each requested chunk, the video that it pertains to, 
its index in that video, the requested bitrate, and 
the download start\fshyp{}end time. Finally, we have developed
a screen analysis tool using pyautogui and 
opencv~\cite{itseez2014theopencv} to record duration 
of each rebuffering event. \S\ref{ss:env} further details our setup.


%



\subsubsection{Chunk Download Control}
\label{ss:tt-algo}



In this section, we introduce TikTok's chunk download ordering and scheduling
which depend both on instantaneous network throughput and the client's
internal buffer status.
Figure~\ref{fig:playback-demo} serves as the foundation of our analysis, 
illustrating the download decisions 
(\emph{i.e.}, order and timing of chunk downloads across videos, 
bitrates used each time) 
that TikTok makes during a 
representative two\hyp{}minute session in which we 
replay a 21\hyp{}swipe user trace. 
We plot client\hyp{}side playback buffer occupancy 
in Figure~\ref{fig:buffer_occupancy}, which shows 
the number of videos with at least one downloaded (but unplayed) chunk.
We see that TikTok spends most of its time downloading the first chunk of 
videos, and downloads the second chunk when and only when the video starts to play, 
\eg, the download of the second chunk of video two and the play\hyp{}start of video two 
start simultaneously at $t=22$ seconds.

Our analysis indicates that TikTok proceeds according to three discrete
states, cycling among the three in order to handle one \emph{group\hyp{}of\hyp{}ten}
videos. When the app starts and at the beginning of every 
group\hyp{}of\hyp{}ten, TikTok starts in a \textbf{ramping\hyp{}up state} 
where it continuously downloads the first chunks to build up the 
buffer as quick as possible.  

After accumulating five first chunks 
at $t=18$ seconds, TikTok starts to play the buffered video 
and at the same time enters the \textbf{maintaining state}, 
where TikTok aims to maintain a constant five first chunks in its buffer.
When TikTok switches to play a new video (due to user swipe or the playback 
reaching the end of a video),
the player fetches one first chunk from the buffer, 
triggering TikTok to immediately initiate the download of 
the first chunk of next video listed in the manifest file, as indicated
by the additional download events corresponding to either swipes
or video changes due to end of video in the green ``maintaining state'' 
regions of Figure~\ref{fig:playback-demo}.
We see in Figure~\ref{fig:buffer_occupancy} that as the downloading of 
each first chunk finishes, buffer levels return to five, high
water mark buffering level TikTok has chosen.

The advantage of being in the maintaining state is that 
in this state, TikTok is quite resilient to quick
user swipes. For example, in the second group\hyp{}of\hyp{}ten 
of Figure~\ref{fig:playback-demo_all} ($t=110$),
the user swipes early in multiple consecutive videos, quickly draining the buffer.
Even under such challenging circumstances, TikTok experiences no rebuffering 
since its buffer contains five first chunks.

Finally, after downloading all the first chunks of the 10 videos 
listed in the current manifest file,
TikTok enters the \textbf{prebuffer\hyp{}idling state},
where TikTok stops initiating any new downloads of 
first chunks.
Meanwhile, TikTok continues video playback, consuming 
video chunks in its buffer,
so buffer occupancy decreases monotonically in this state, as shown in Figure~\ref{fig:buffer_occupancy}.
Our hypothetical explanation of this idle period is that 
TikTok is waiting to measure the user's reaction 
(swiping early means they might not be interested in the content) 
to the videos TikTok recommends in last round (manifest file),
so it can assess its recommendation quality 
and adjust the subsequent round's recommendation 
before sending the next manifest file.

In contrast to the resilience of the maintaining state,
TikTok becomes somewhat vulnerable in the prebuffer\hyp{}idling state, 
where TikTok drains the buffer by itself. 
For example, TikTok experiences rebuffering in the middle of 
two video groups in Figure~\ref{fig:playback-demo_all}.
At that moment, TikTok has no buffered first chunk and 
at the same time spends a long time downloading the second chunk of the current video,
leaving no time budget for downloading the first chunk of next video.
In such a case, one user swipe results in rebuffering.

When the user watches nine out of 10 videos listed in the current manifest file, 
TikTok exits the prebuffer\hyp{}idling state and begins afresh
in the ramp\hyp{}up state,
to download the videos listed in the next manifest file.
We observe from Figure~\ref{fig:playback-demo_all} that 
TikTok repeats these three states every group\hyp{}of\hyp{}ten
videos listed in the manifest file. 

\subsubsection{Network and Swipe Input Adaptation} 

We now investigate the effects of swipes, buffer occupancy, and the network 
on TikTok's bitrate and buffering choices.

\begin{figure}
    \centering
    \begin{subfigure}[t]{0.535\linewidth}
      \centering
      \includegraphics[width=\linewidth]{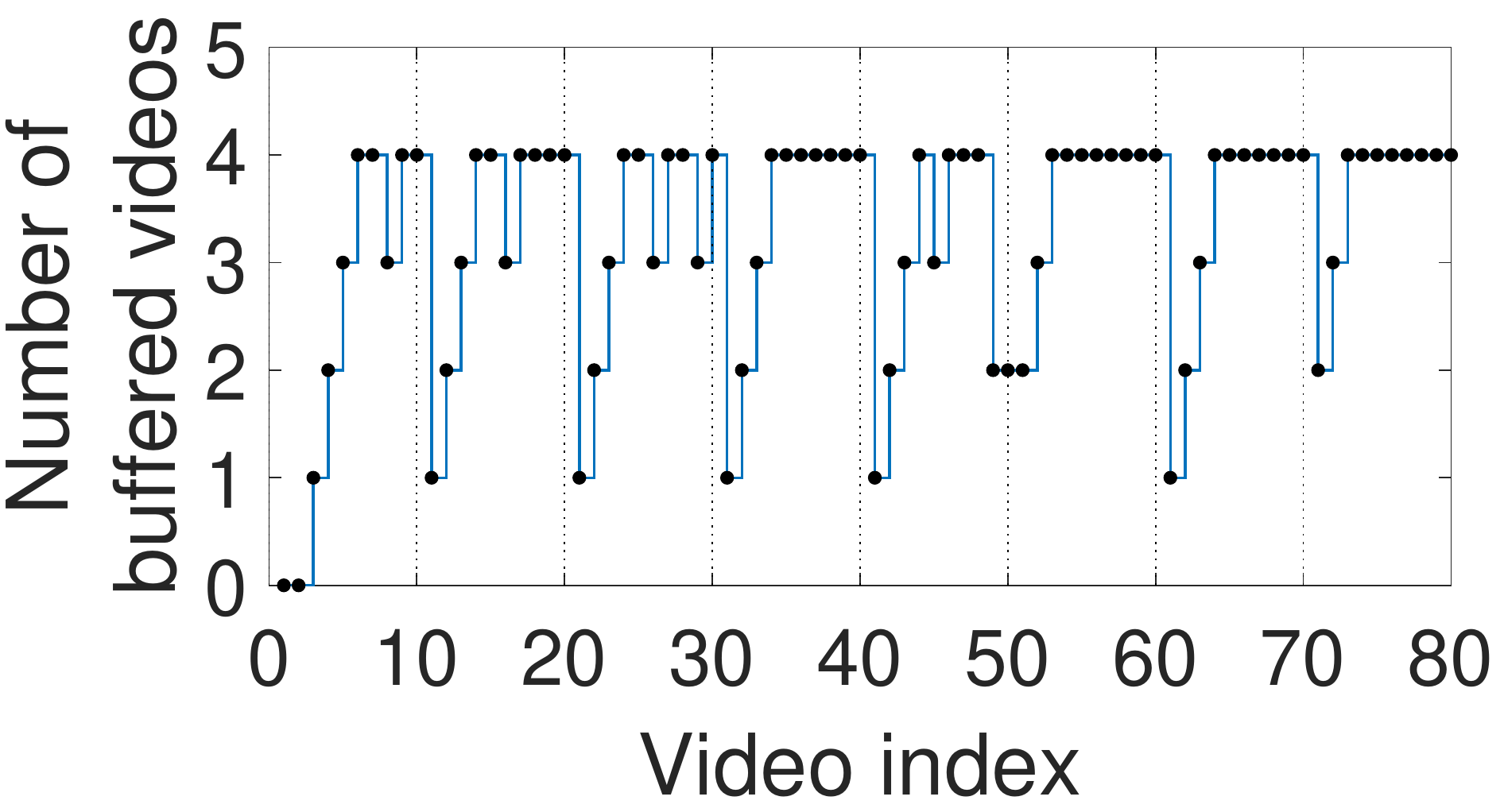} 
      \caption{Network throughput 10~Mbit/s.}
      \label{fig:10mbps}
  \end{subfigure}\hfill
  \begin{subfigure}[t]{0.45\linewidth}
      \centering
      \includegraphics[width=\linewidth]{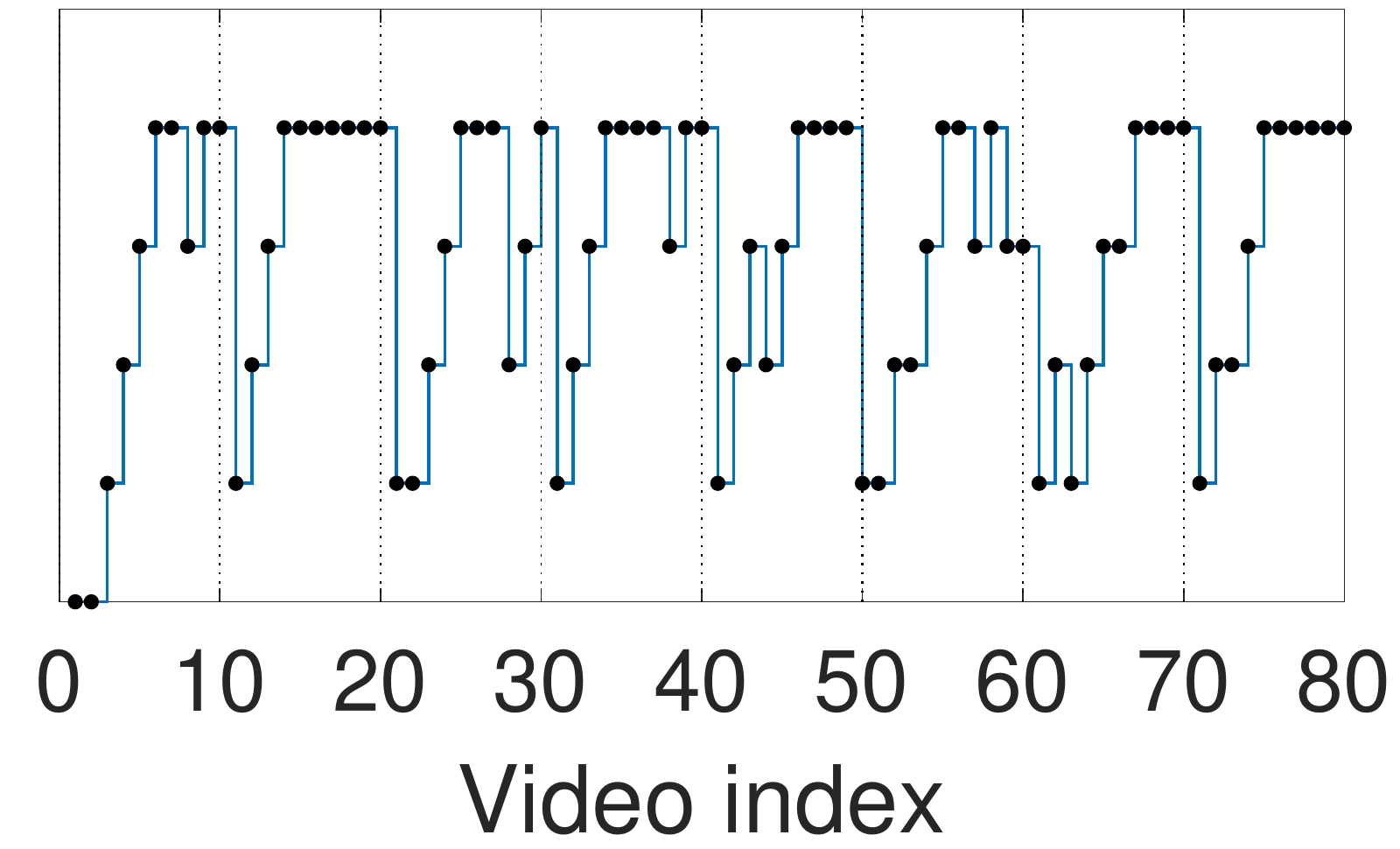} 
      \caption{Network throughput 3~Mbit/s.}
      \label{fig:3mbps}
  \end{subfigure}\hfill
    \caption{
    The number of downloaded videos inside the buffer when TikTok starts to download the first chunk of a new video 
    via networks with capacity of \textbf{(a)} 10~Mbit/s and \textbf{(b)} 3~Mbit/s.}
    \label{fig:buffer_download}
    \vspace{-10pt}
\end{figure}

To measure the impact of network throughput on the TikTok's buffering strategy, 
we conduct a controlled experiment by setting the network 
capacity to 10 and 3~Mbit/s using Mahimahi.
We plot the number of buffered first chunks at the moment 
TikTok initiates a downloading of the first chunks
in Figure~\ref{fig:buffer_download}.
Combing Figure~\ref{fig:10mbps} and~\ref{fig:3mbps}, 
we see that TikTok adopts the same buffering strategy even when the network capacity decreases. 
As we will introduce in the following section that the network throughput mainly affects
the selected bitrate not the buffering strategy.

\begin{figure}
    \centering
    \includegraphics[width=0.8\linewidth]{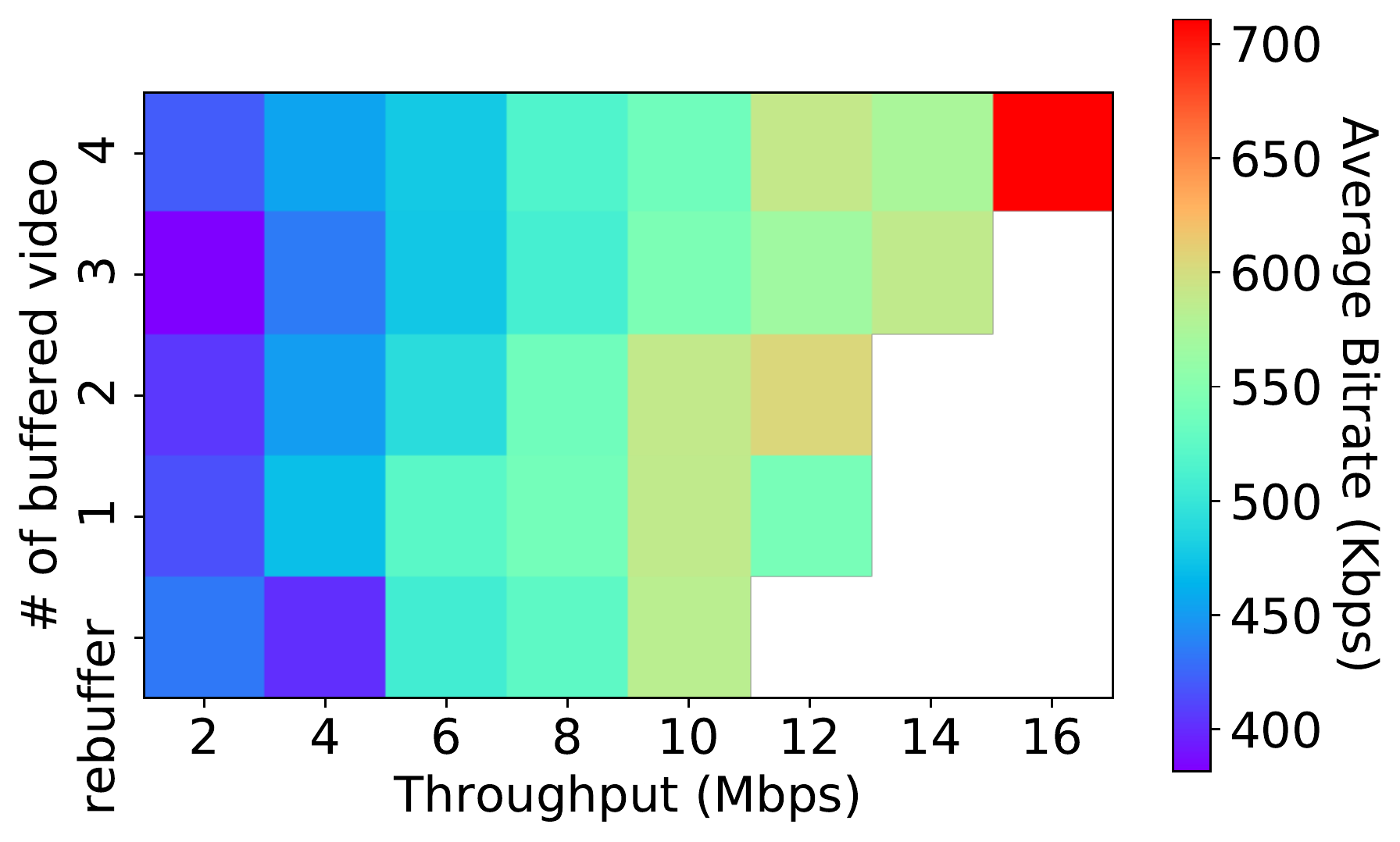}
    \caption{Impact of instantaneous network throughput and client 
    video buffer occupancy on TikTok's chosen video bitrate.}
    \label{fig:bitrate-analysis}
    \vspace{-10pt}
\end{figure}

Next, we analyze the joint impact of network throughput and buffering
status on the bitrate decisions TikTok makes.  We 
collect instantaneous network throughput and buffer 
status coupled with TikTok's bitrate decisions, for 5,300 videos,
and plot the results in Figure~\ref{fig:bitrate-analysis}.
In the figure, the x-axis is the network throughput of the one-second 
period before the downloading of that video,
\ie, the time period within which TikTok makes its decisions about the bitrate.
The y-axis is the number of downloaded first chunks in the buffer.
The color of a tile represents the average bitrate $R$ of the video,
which is given by $R = S/L$ where $S$ is the size of the video in
bits and $L$ is the length of the video in seconds.
Some tiles are not colored because the combination of the throughput and buffer status 
is not observed during our measurement, 
\eg, when the throughput is 16~Mbit/s, we always observe four 
downloaded first chunks in the buffer. 
We observe that bitrate decisions 
correlate positively with network throughput, but
observe no evidence for correlation with buffer status.

\subsubsection{Limitations of Current Short Video Streaming}


Despite pre\hyp{}buffering the beginnings of short videos, 
TikTok has a fundamentally static approach 
to coping with the uncertainty of future user swipes, 
with no adaptation across different videos or users. 
This static approach is often too cautious or aggressive, 
manifesting in two particular ways, with the intended 
goal of minimizing issues with fast swipes:

\parahead{Lack of swipe prediction.} 
TikTok prioritizes the downloading of the first chunk, assuming 
that the user always swipes frequently, 
and delays the downloading of the second chunk to the beginning of the video playing.  
As we will show in the next section however,
there are indeed some users swipe early when watching a video, 
but there also exists a significant number
of users who watch most of many videos and swipe at the end or not at all.
So, the urgency of downloading the second varies with users and videos: 
a fixed rule cannot handle all the cases. 

\parahead{Premature bitrate binding.} TikTok groups the 
first 1~MB data of each video into the first chunk, which 
allows for some certainty in chunk download completion times,
but TikTok selects the bitrate for both chunks 
according to the network conditions
present during the first chunk,
prematurely binding the system into that fixed bitrate 
for the both the first and second chunks.
By design, there is often a large time lag 
between the downloading of the 
first and second chunks of a video, as shown in Figure~\ref{fig:playback-demo},
which our comprehensive measurements confirm 
(the median gap between
first and second chunk downloads is 25~s., with
an interquartile range of 23~s.),
resulting in a potential mismatch with 
network throughput conditions that change
in this interim time period.

\paragraph{Network Idling.} As shown in Figure~\ref{fig:playback-demo}, TikTok leaves a significant among of time idling without downloading video content. In the contrast, the bitrate of videos still has room to improve. This also calls for a better ABR algorithm to stream higher bitrate video by utilizing the idle time in a better way.

%% file: user_swipes.tex
\section{Characterizing User Swipes}
\label{s:measurement-swipe}

\begin{figure}
\centering
      \includegraphics[width=0.7\linewidth]{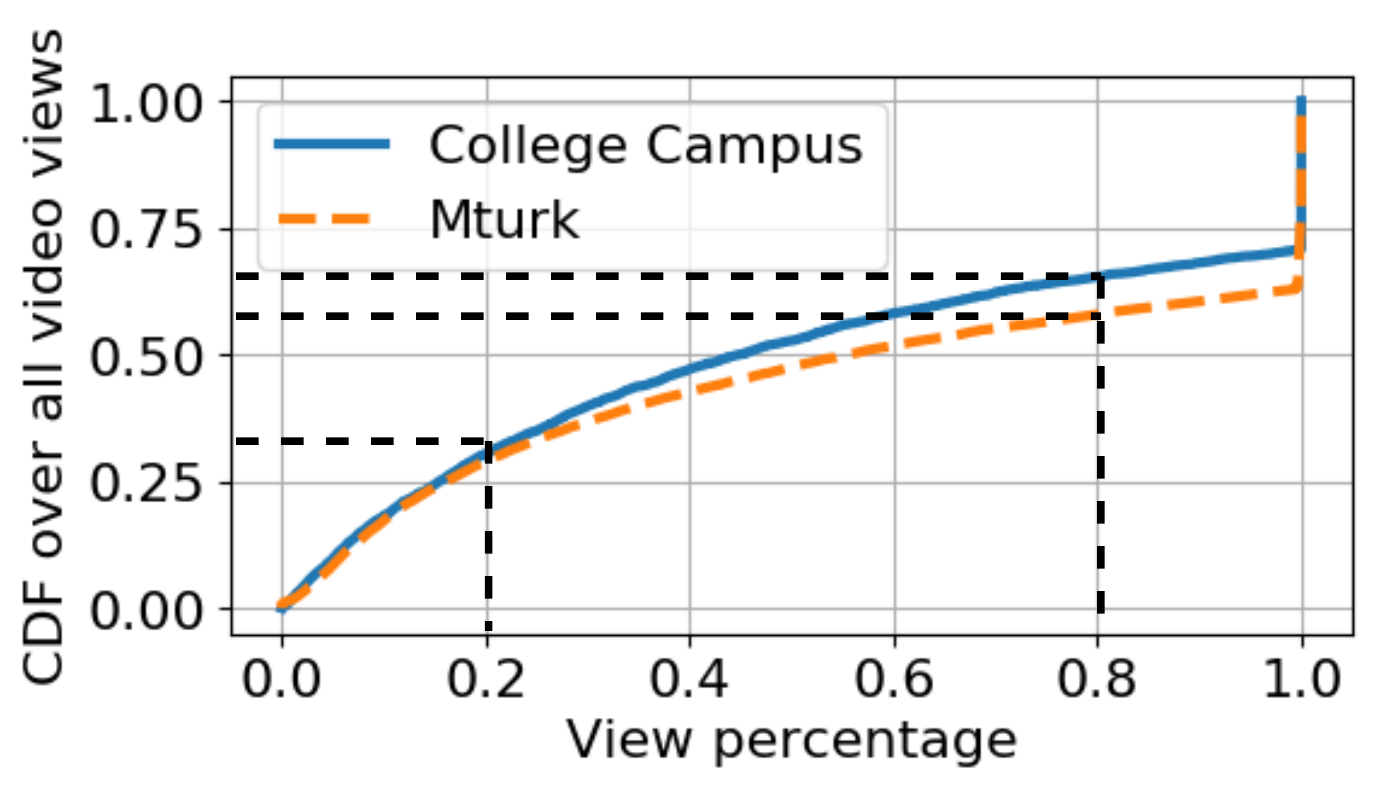} 
      \caption{The distribution of average viewing percentage across all short video views.}
      \label{fig:swipe-total}
  \end{figure}

\begin{figure}
    \centering
    \includegraphics[width=\linewidth]{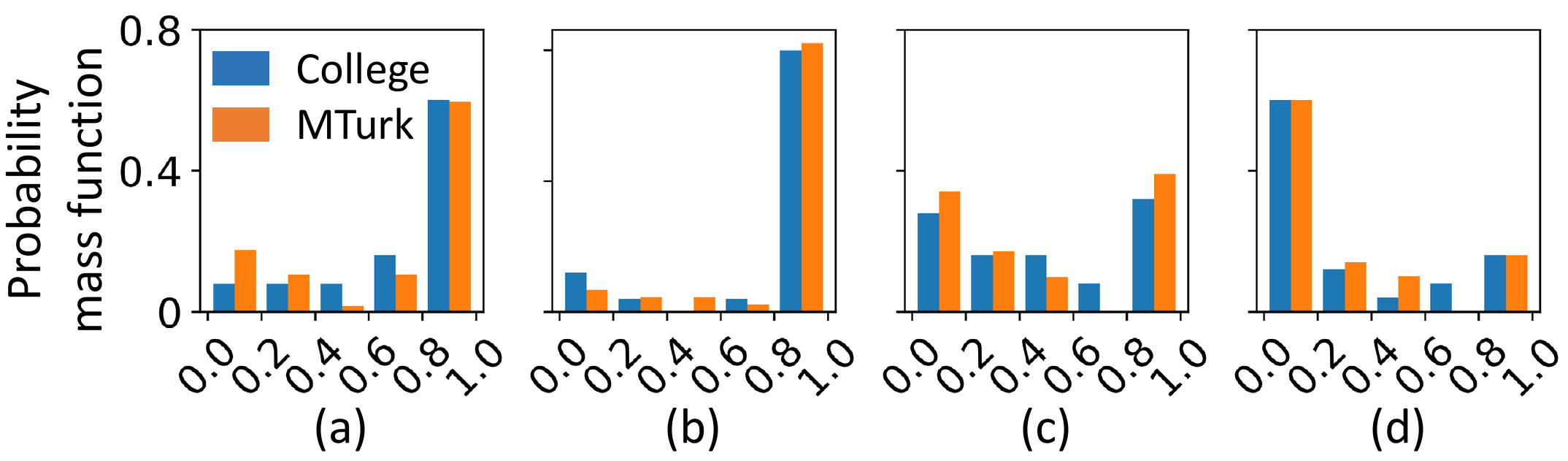}
    \caption{Distribution (across different users) of video viewing percentage for four sample videos \emph{(a)}--\emph{(d)}.}
    \label{fig:swipe-by-video-example}
\end{figure}

To better understand the nature of user interactions 
(\emph{i.e.}, swipes) with short video applications, we 
conducted two IRB\hyp{}approved user studies. In each study, 
we present users with a web-based short video streaming 
service that resembles the interface offered by TikTok 
We considered 500 popular short videos gathered by crawling
the videos displayed on the TikTok landing page over time.
The videos were randomly ordered per session, and each 
user watches 20 minutes of video with the ability to 
swipe freely (all swipes are recorded). Note that the 
number of videos watched by a given user depends on 
the number of swipes they performed.

\parabreak{}For generality, we performed two 
versions of this study:

\parahead{1.~College campus:} we recruit 25 student volunteers who collectively swipe 3,069$\times$ across the study.

\parahead{2.~Amazon Mechanical Turk (``MTurk''):} we recruit 
258 different users. To ensure active user participation,
we augment our web application to inject random 
interactivity tests that ask users to swipe within 10 
seconds. Users who 
fail to swipe in time are entirely excluded from the study; 
users who do swipe continue the experience, but we exclude
the forced swipe(s) from our final dataset. In total, we
retain data from 133 workers, which covers 15,344 swipes.

\begin{figure*}
    \begin{minipage}[t]{0.33\linewidth}
        \centering
        \includegraphics[width=\linewidth]{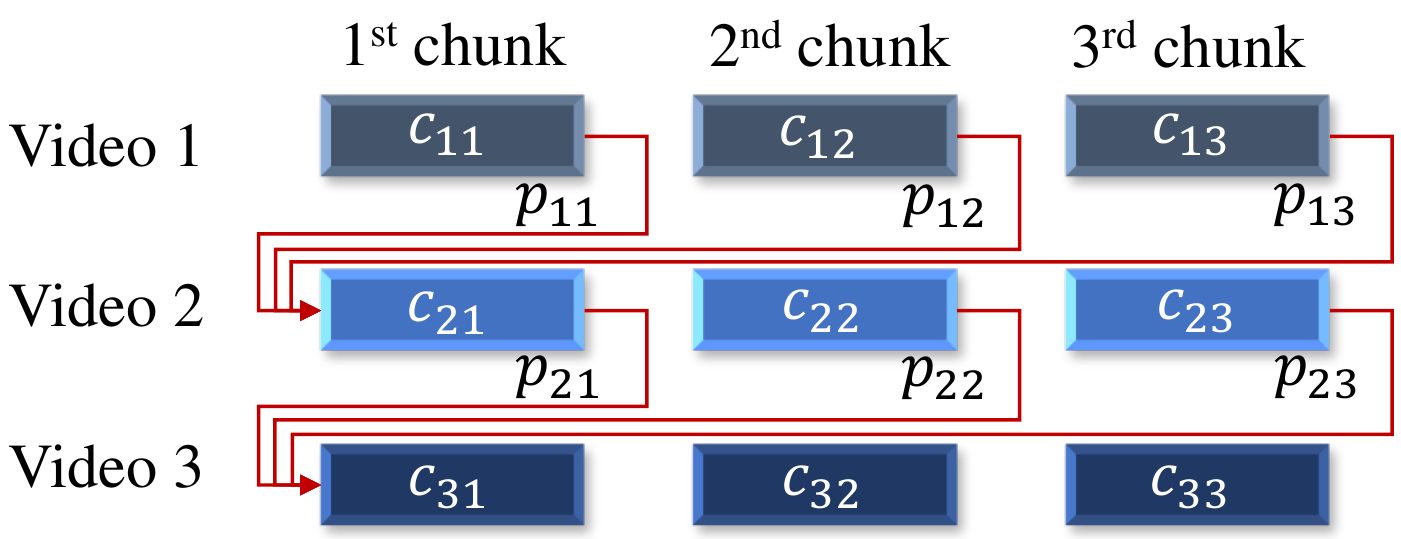} 
        \caption{Short video streaming model:
        the player plays videos sequentially, 
        switching to the first chunk of the next video after a swipe.
        }
        \label{fig:system_model}    
    \end{minipage}\hfill
    \begin{minipage}[t]{0.33\linewidth}
        \centering
        \includegraphics[width=\linewidth]{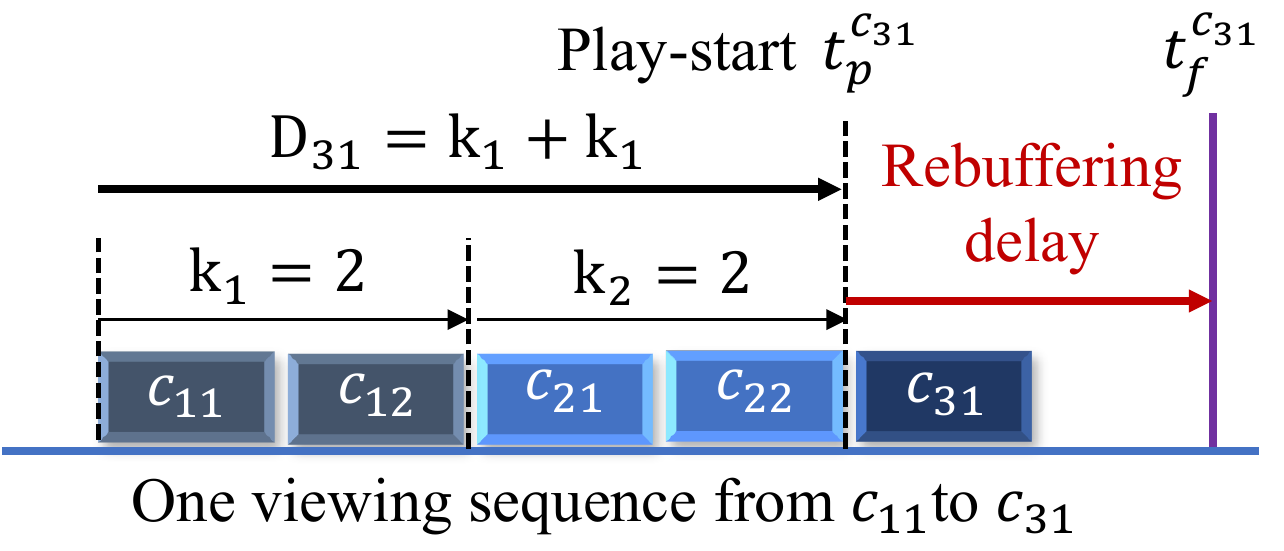} 
        \caption{Chunk rebuffering delay depends on the order 
        between the play start time $t_p$  
        and the download finish time $t_f$.
        }
        \label{fig:exp_rebuffer}    
    \end{minipage}\hfill
    \begin{minipage}[t]{0.31\linewidth}
        \centering
        \includegraphics[width=\linewidth]{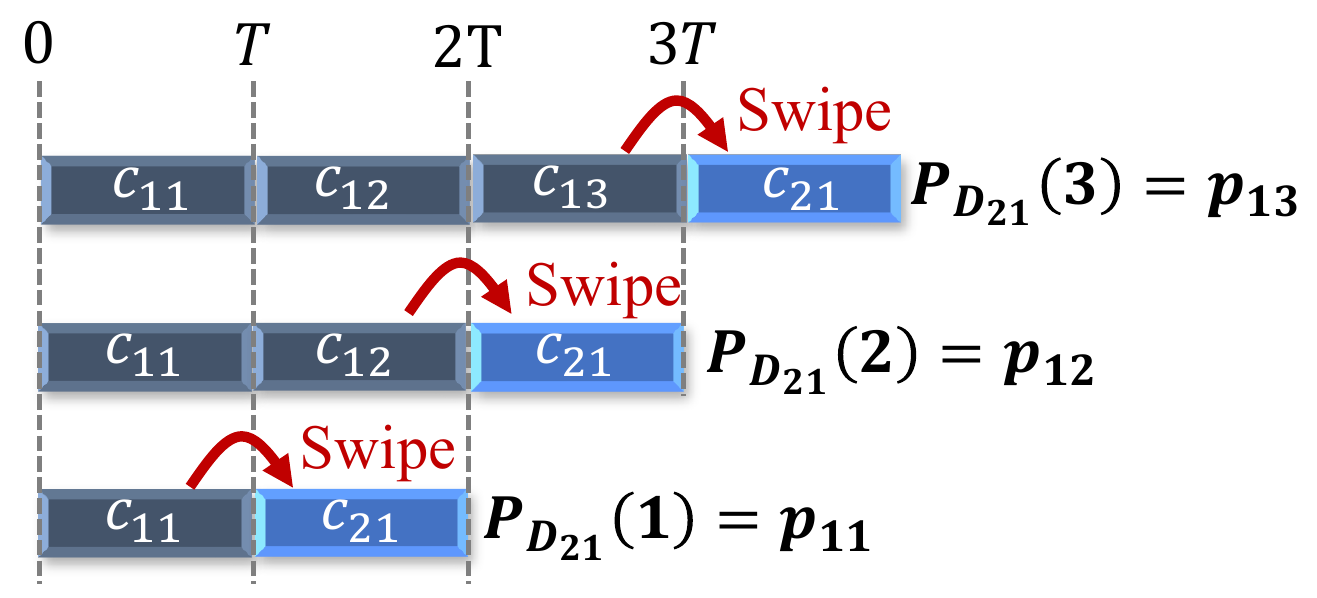} 
        \caption{Three possible viewing sequences that start from chunk $c_{11}$ and end at chunk $c_{21}$.}
        \label{fig:3seq}
    \end{minipage}
\end{figure*}

\paragraph{Overall swipe distributions.}
Figure~\ref{fig:swipe-total} shows the distribution
of swipe times across all video-user pairs in both 
studies. As shown, users are most 
likely to swipe either soon after the video playback begins 
or at the tail end of the video (manually or via auto-swiping 
once the video completes); this is consistent with prior studies
on user swipe patterns~\cite{zhang2022measurement}. For 
instance, 29\% and 42\% of swipes from MTurk users are 
within the first 20\% or last 20\% of videos, respectively. 
Swipes between these two endpoints occur, but far less 
often and with increasingly low likelihood as users 
watch more videos, \emph{e.g.}, only 6\% of swipes in the College 
Campus dataset are in the 60--80\% of videos. 

\paragraph{Swipe distributions per video.} Figure~\ref{fig:swipe-by-video-example} shows the swipe probabilities for four representative videos, aggregated across the users who watched each one in the two studies. As shown, different videos can exhibit significantly different patterns in terms of user swipe probabilities. For instance, over 60\% of swipes in videos (a) and 80\% of swipes in videos (d) come within the last few seconds of the videos (indicating low swipe probabilities for those videos). Video (c) exhibits the opposite pattern---60\% of swipes in the first 20\% of the video (indicating high swipe probabilities)---while swipes in (b) are more evenly distributed across the start and end of the video. Perhaps more importantly, we observe substantial stability in the swipe distributions per video across different user datasets: KL divergence values between the MTurk and College Campus datasets are 0.2 and 0.8 for the median and 95th percentile videos, respectively.

\paragraph{Takeaway.} Despite general similarities in their swipe patterns, users follow a few different modes of swiping (e.g., swiping early in the chunk vs. not at all), each of which warrants a different buffering strategy to ensure high QoE. Fortunately, cross-user swipe data that is aggregated \emph{per video} provides a relatively stable indicator as to how likely swipes are (and will be) in a given video, and (more coarsely) at what part of the video they will occur. We show in \S\ref{s:design} how \systemname\ leverages this coarse information -- which is readily available at existing short video servers -- to make robust buffering decisions that handle cross-user swipe traces.

%% file: Design.tex
\section{Design}
\label{s:design}
The core idea behind \systemname{} is to leverage the stability in swipe distributions across videos (\S\ref{s:measurement-swipe}) to get 
a \emph{coarse} sense of the likelihood of swipes at different video chunks. 
Coupling this information with constraints on inter-chunk viewing 
sequences intrinsic to short video applications, 
\systemname{} models the \emph{expected rebuffering time} for 
each potential chunk as a continuous function over the expected 
download and playback times (\S\ref{ss:expected_rebuf}). 
\systemname{} then employs a greedy algorithm atop those 
functions to find a chunk buffering sequence that 
minimizes the expected rebuffering delay over a time horizon for 
a given network throughput estimate, and across different 
user viewing sequences. Lastly, 
\systemname{} feeds that buffering sequence into a
bitrate selection algorithm (RobustMPC~\cite{MPC} in our 
implementation) that determines the bitrate to use per 
chunk to optimize overall QoE (\S\ref{ss:buffer_order}).



\subsection{Forecasting Rebuffering Delay}
\label{ss:expected_rebuf}

\systemname{}'s expected rebuffering functions aim to quantify
user-perceived delays across different chunk download times and viewing sequences. For clarity of exposition, we begin by explaining the construction of these functions in a discrete setting where users can only swipe at chunk boundaries; we then extend the discussion to incorporate continuous functions and arbitrarily-timed user swipes.

\parahead{System Model.}
Short video apps follow the flow shown in Figure~\ref{fig:system_model}.
Each video consists of multiple chunks of \emph{chunk time} $T$.
Within the $i^{\mathrm{th}}$ video with $N_i$ chunks, if the user does not
swipe, the video player plays its chunks $c_{ij}$ sequentially,
where $j \in [0, N_i]$ is the chunk index.
When playback reaches the end of the video or the user swipes,
the player jumps to the first chunk of the next video.
Since user swipe distributions vary across videos
(\S\ref{s:measurement-swipe}),
we denote the probability that the 
user swipes after watching chunk $c_{ij}$ as $p_{ij}$. 
The list of the chunks the user watches 
is a \emph{viewing sequence} 
\begin{equation}
   V_s = [ c_{11},\ldots, c_{1k_i}, c_{21},\ldots, c_{K1},\ldots, c_{Kk_L} ] \label{eqn:vs}
\end{equation}
where the user views the first $k_i$ chunks of 
the $i^{\mathrm{th}}$ video, assuming that the user
watches $L$ videos in total.
Then the probability distribution of $k_i$ is $P_{k_i} = \{p_{i1}, p_{i2}, \ldots ,p_{iN_i}\}$. We define $D_{ij}$, the number of 
chunks that a user has watched prior to chunk $c_{ij}$:
\begin{equation}
   D_{ij} = \sum_l^{i-1} k_l + (j-1).  
   \label{eqn:d}
\end{equation}
By knowing the number of chunks that a user has watched 
before $c_{ij}$, the playback start time of
$c_{ij}$ then will be $D_{ij}\cdot T$.
As shown in Figure~\ref{fig:exp_rebuffer}, the expected 
rebuffering delay for some chunk $c$ depends 
on the relationship between the chunk's \textit{play start} 
time $t_p^c$ and \textit{download finish} time $t_f^c$. 
There exists no rebuffering if the chunk downloading finishes before the play start time.
Otherwise, rebuffering happens and 
the time difference between $t_p^c$ and $t_f^c$ 
tells us $c$'s \emph{rebuffering delay}:
\begin{equation}
    T^{\mathrm{rebuf}}_{c}(t_f^c, t_p^c) =\left\{
        \begin{array}{cc}
             0,          &  t_f^c < t_p^c \\
             t_f^c - t_p^c, &  t_f^c \ge t_p^c
        \end{array}
        \right.
        \label{eqn:t_rebf_single}
\end{equation}
The play start time of each chunk is determined by the viewing sequence $V_{s}$,
as shown in Figure~\ref{fig:exp_rebuffer}.
Since our goal is to schedule $c$'s download to minimize 
rebuffering, we now formulate a reward function to meet this goal, parameterized on $t_f^c$ and averaging over all 
possible viewing sequences (which are not under our control).
The \emph{expected 
rebuffering delay} of chunk $c$ given that chunk's download
finish time $t_f^c$, across all possible viewing sequences, is:
\begin{equation}
     \mathbf{E}^{\mathrm{rebuf}}_{c}(t_f^c)  =  \sum_{V_{s} \in \Phi} \Pr(V_{s}) \cdot T^{\mathrm{rebuf}}_{c}(t_f^c, t_p^c(V_s)) 
\label{eqn:e_rebf}
\end{equation}
where probability $\Pr(V_{s})$ represents 
how likely a specific viewing sequence $V_{s}$ will appear 
based on user swipe distribution data,
$t_p^c(V_s)$ is $c$'s play start time in $V_s$, 
and $\Phi$ is the set of all possible viewing sequences. 

\label{s:per_chunk_rebf}

To calculate the expected rebuffering delay for a specific chunk,
we need to enumerate all possible viewing sequences that reach 
this chunk, as Eq.~\ref{eqn:e_rebf} shows.
For each sequence, we need to compute how likely 
this sequence will appear based on the user swipe probability and 
then determine the play start time of that specific chunk. 
Based on short video chunk playback constraints (\S\ref{s:intro}, p.~\pageref{s:intro}), we propose separate algorithms for calculating the expected rebuffering delay of a video's first chunk, and remaining chunks,
respectively.




\paragraph{First chunk of a video.}
The number of possible viewing sequences between 
chunk one of video one ($c_{11}$)
and chunk one of video $i$ ($c_{i1}$) increases 
exponentially with $i$. On the other hand, the 
number of sequences from the first chunk 
of the previous video to the first chunk of the current video is 
bounded by the number of chunks in the former.
For example, as shown in Figure~\ref{fig:system_model},
there are only three possible viewing sequences from 
chunk $c_{21}$ to chunk $c_{31}$.
We therefore enumerate the viewing sequences in a recursive manner:
deriving the viewing sequences that reach the first chunk 
of the $i^{\mathrm{th}}$ video 
based on the viewing sequence of the first chunk of the 
$(i-1)^{\mathrm{st}}$ video.

\begin{figure}[htb]
    \centering
    \includegraphics[width=\linewidth]{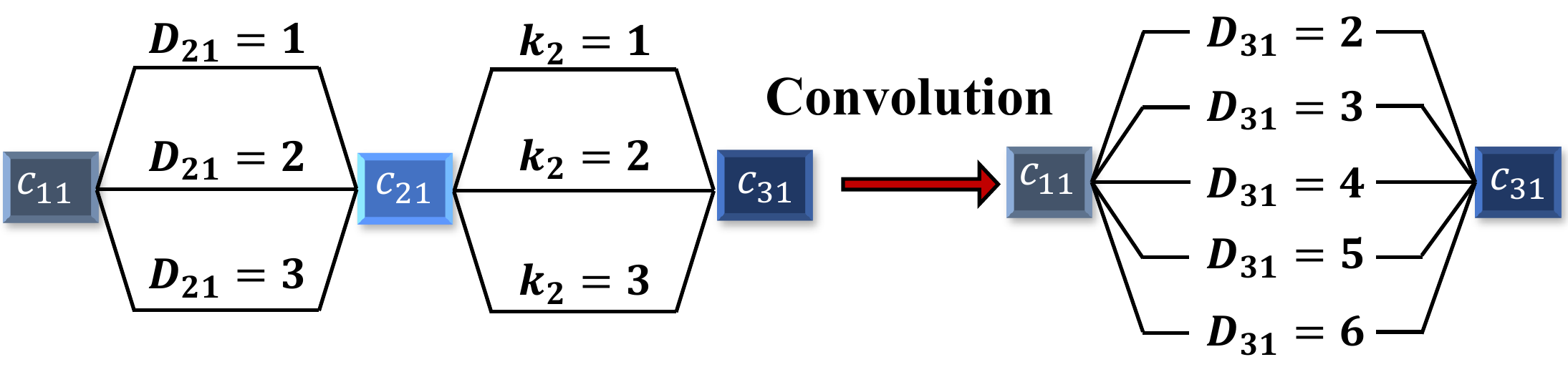} 
    \caption{
    Convolution of the $P_{D_{21}}[\cdot]$ and $P_{K_2}[\cdot]$ provides us the probability distribution $P_{D_{31}}[\cdot]$.}
    \label{fig:merge}
\end{figure}

We start from the base case, viewing sequences 
from $c_{11}$ to $c_{21}$.
Figure~\ref{fig:3seq} lists all three possible viewing sequences
that start from $c_{11}$: we see that random variable
$D_{21} = k_1$ (\emph{cf.}~Eq.~\ref{eqn:d}). Similarly, as shown in Figure~\ref{fig:exp_rebuffer}, $D_{31} = D_{21} + k_2$
(\emph{cf.}~Eq.~\ref{eqn:vs}).
The distribution of random variable $D_{31}$ can then 
be calculated as: 
\begin{equation}
    P_{D_{31}}[n_0] = \sum_{i=1}^{n_0-1} P_{D_{21}}[i] \cdot P_{k_2}[n_0-i]
\end{equation}
where $P_{D_{31}}[n_0]$ means the probability of there are $n_0$ chunks before chunk $c_{31}$ is viewed. 
This formula by definition is the operation of convolution between $D_{21}$ and $k_2$,
as shown in Figure~\ref{fig:merge}.
Without losing generality, the number of chunks that user watches  before chunk $c_{i1}$, $D_{i1}$, is the sum of $D_{(i-1)1}$ and $k_{i-1}$. Therefore the distribution of $D_{i1}$ is the convolution of $D_{(i-1)1}$ and $k_{i-1}$:
\begin{equation}
    P_{D_{i1}} = P_{D_{(i-1)1}}\ast  P_{k_{i-1}}
\end{equation}
With the knowledge of $D_{i1}$'s distribution for the first chunk of all the videos,  we calculate the expected rebuffering delay of chunk $c_{i1}$, as the function of download finish time:
\begin{equation}
     \mathbf{E}^{\mathrm{rebuf}}_{c_{i1}}(t_f)  =  \sum P_{D_{ij}}[n] \cdot T^{\mathrm{rebuf}}_{c_{i1}}(t_f, (n+1)T) 
\label{eqn:e_rebf_31}
\end{equation}

\begin{figure}[!t]
    \centering
    \includegraphics[width=0.9\linewidth]{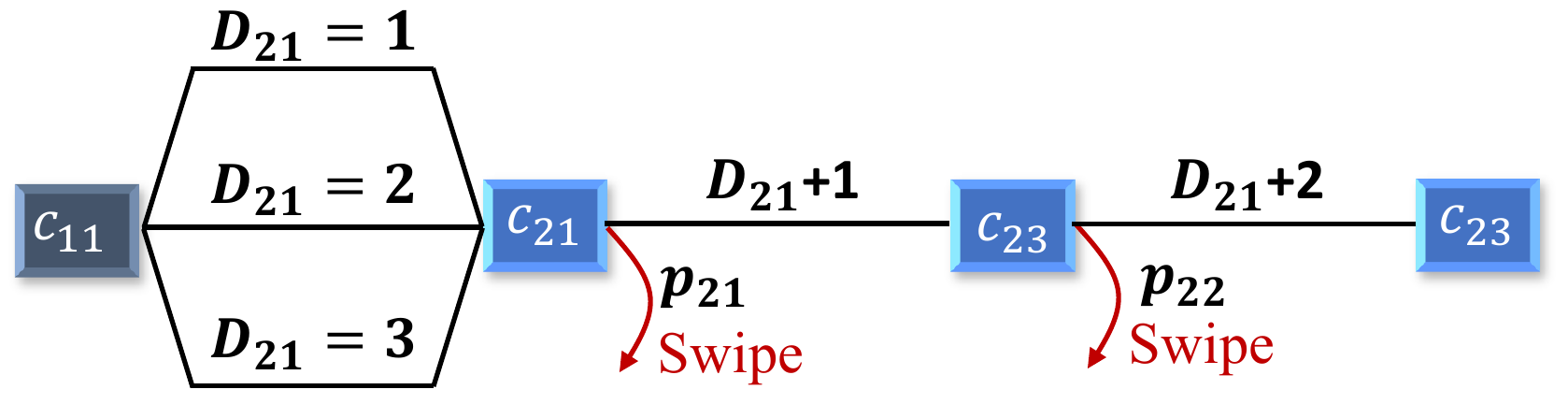} 
    \caption{
    Starting from chunk $c_{21}$, the user must watch
    the second video continuously with swiping to reach chunk $c_{23}$.}
    \label{fig:seq_merge_2nd}
\end{figure}

\paragraph{Remaining chunks in a video.}
There exist only one possible viewing sequence from the first chunk to later chunks in the same video.
For example, as shown in Figure~\ref{fig:seq_merge_2nd},
the video chunk $c_{23}$ will be played when and only when the user watches the $i=2^{\mathrm{nd}}$ video continuously without swiping. For non-first chunk $c_{ij}$, the number of chunks that user watched before it, $D_{ij}$, is the summation between the $D_{i1}$ and $j-1$ since the user has to watch the first $j-1$ chunks in video~$i$ before starting watch it. Then the distribution of $D_{ij}$ should be the distribution of $D_{i1}$ delayed by $j-1$ chunks. In addition, the user might swipe to the next video before watching $c_{ij}$. To summarize this, we can calculate the distribution of $D_{ij}$ for non-first chunk as:

\begin{equation}
    P_{D_{ij}}[n_0] = P_{D_{i1}}[n_0-(j-1)] \times (1 - \sum_{m=1}^{j-1}p_{im})
\end{equation}
With the distribution of $D_{ij}$, we follow the same procedure
to calculate expected rebuffering time for remaining chunks
in a video, according to Eq.~\ref{eqn:e_rebf_31}.

\paragraph{Arbitrary user swipes.}
In reality, the users' swipe does not only happen after a chunk 
finishes. We extend our analysis to allow the user to swipe 
at arbitrary time points when playing $i^{\mathrm{th}}$ video.
If the user (continuously\hyp{}valued) viewing time for 
video~$i$ is $\kappa_i$, 
the PDF of $\kappa_i$ is $f_{\kappa_i}(t_0)$. 
The \emph{play start time} of chunk $c_{ij}$, $\Delta_{ij}$,
is a random variable, with probability density 
function $f_{\Delta_{ij}}(t)$. For the first 
chunk of video $i$, its playing start time $t_f^{c_{i1}}$ is also the summation of the playing start time of the previous video $t_f^{c_{(i-1)1}}$ and the time the user spends watching the previous 
video $\kappa_{i-1}$. Following a similar principle, we compute $f_{\Delta_{i1}}(t)$ for the first chunk of a video $i$ as
\begin{equation}
f_{\Delta_{i1}}(t) = f_{\Delta_{(i-1)1}}(t) \ast  f_{\kappa_{i-1}}(t).
\end{equation}
For subsequent chunks $c_{ij}$, we also calculate the
playing start distribution based on the first chunk
\begin{equation}
f_{\Delta_{ij}}(t) = f_{\Delta_{(i-1)i}}(t - (j-1)\cdot L)
 \cdot \left(1 - \int_{0}^{(j-1)\cdot L} f_{\kappa_{i}}(x)\right) dx
\end{equation}
Then the expected rebuffering function can be calculated similarly to Eq.~\ref{eqn:e_rebf_31}:
\begin{equation}
     \mathbf{E}^{rebf}_{c_{ij}}(x)  =  \int_{t=0}^{x} f_{\Delta_{ij}}(t) \times T^{rebf}_{c_{ij}}(x, t) dt
\label{eqn:e_rebf_cont}
\end{equation}

\subsection{Determining Buffering Sequences}
\label{ss:buffer_order}

Given the preceding computation of expected rebuffering 
delay for each chunk, \systemname{}'s next task is to determine
an order of chunks to download (\emph{i.e.}, a buffering 
sequence) that minimizes expected rebuffering delay
over a lookahead horizon. Prior schemes (\emph{e.g.}, 
MPC \cite{MPC}) can then be used to determine the bitrates for
those chunks to optimize overall QoE for the horizon. 
However, unlike prior schemes, the horizon that we use is
based on time (not chunks). The reason is that different 
user swipe patterns can translate into different numbers 
of viewed chunks---using a horizon sized to a fixed number 
of chunks could result in optimization over very short viewing 
times (negating the effects of longer-term planning). Our 
current implementation uses a lookahead window of 25 seconds 
based on empirical observations, which is equivalent to the 
five chunks MPC uses.
Intuitively, coarse-grained swipe information suffice to determine
download order for video chunks.  
Chunk ordering relies primarily on whether the user swipes at
the beginning of the video or not: \eg if the user is highly likely 
to not swipe in chunk $c_{11}$, the algorithm then needs to prioritize 
chunk $c_{12}$ over chunk $c_{21}$.


\begin{figure}[!t]
\centering
    \includegraphics[width=\linewidth]{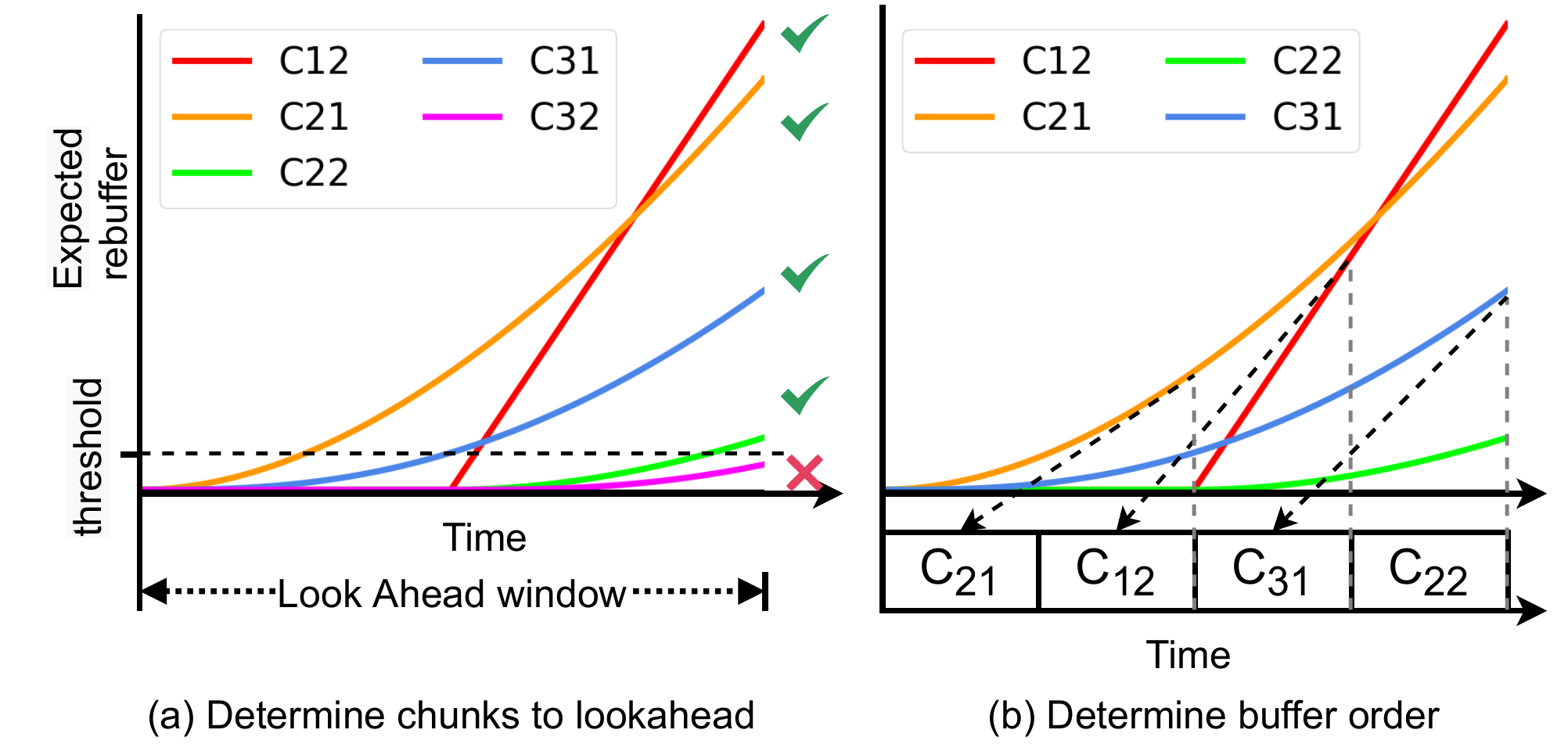}
    \caption{An example to illustrate \systemname's algorithm.}
    \label{fig:algo-demo}
    \vspace{-10pt}
\end{figure}

\subsubsection{Selecting the candidate chunk set}

To determine the set of chunks to consider, we enforce a threshold on the minimum rebuffering penalty that each chunk is expected to incur at the end of the horizon if it is not included in the buffer sequence (Figure~\ref{fig:algo-demo}(b)). Chunks whose rebuffering penalty falls below the threshold are deemed as unlikely to be viewed during the horizon ($C_{32}$ in Figure~\ref{fig:algo-demo}(a)), and thus low priority for inclusion in the buffer sequence. Note that buffer sequences are constructed each time a chunk download completes, so an excluded chunk for one horizon may still be downloaded shortly (via inclusion in the next horizon's buffer sequence). We use an empirically-configured value of $1/\mu$ for threshold, which is the inverse of the rebuffering penalty weight in our target QoE function. 

Using the set of chunks to consider, our final task is to order them in a manner that minimizes expected rebuffering penalties. Our approach is to assign a bitrate to each chunk, and then use an estimate of the network bandwidth to determine when a chunk will complete donwloading (assuming some start time); this, in turn, allows us to compute the expected rebuffering time per chunk using the functions from \S\ref{ss:expected_rebuf}. However, to do this with low computational complexity, since download decisions must be fast, we temporarily assume an equal bitrate per chunk that is set to the maximum bitrate which ensures that all chunks in the list will complete downloading before the horizon completes. Although exclusion of per-chunk bitrate decisions here can result in suboptimal orderings, we find these effects to be minimal (as evidenced by \systemname{} closeness to the optimal scheme in \S\ref{ss:qoe-result}). The reason is that, as discussed in \S\ref{ss:expected_rebuf}, priorities between chunks (and potential per-chunk viewing times) are largely dictated by viewing constraints imposed by the application. Thus, minor discrepancies in bitrates across chunks are thus unlikely to flip the priority order among them.

\subsubsection{Priority-ordering the buffer sequence}

To sort our list of chunks into a buffer sequence, we follow a greedy algorithm, whereby we partition the horizon into chunk-sized slots. For a given slot $i$, we select the chunk that will incur the largest additional rebuffering penalty if it were to be scheduled in slot $i+1$ rather than $i$. Figure~\ref{fig:algo-demo}(b) shows this process for a scenario in which chunk $c_{11}$ just completed downloading: $c_{21}$ is assigned to slot 1 as its rebuffering penalty jumps the most between slots 1 and 2; $c_{12}$ is next as it has the highest penalty for not going in slot 2, and so on.
Finally, using the generated buffer sequence, \systemname{} applies MPC's algorithm to determine the bitrate for each chunk in the buffer sequence in a way that optimizes the entire QoE (not just minimizing rebuffering) for the horizon according to the forecasted network throughput, \emph{i.e.}, the harmonic mean over the observed throughputs in the last 5 chunk downloads.

\section{Implementation}
\label{ss:implementation}
\systemname{} makes no change to the CDN\fshyp{}server
side so our system can be easily deployed 
client side.
\systemname{} includes one control module and multiple buffer 
modules. Each buffer module manages the video playback of 
one short video, including downloading chunks, 
tracking playback progress, and reporting buffer status. 
We reuse the DASH.js playback management 
for the buffer modules. The control module manages scheduling across short videos, collecting estimated throughput and buffer length 
from each buffer module. With the collected data, control module runs 
\systemname's algorithm to schedule the video buffering. 
Based on the algorithm's output, it assigns the 
quota to the buffer module that is assigned to download 
the next video chunk. 
The quota includes the target video bitrate and the 
target download finish time. 
Once the buffer module receives the quota, it sends an 
HTTP request with target bitrate to the CDN to download 
the corresponding video chunk. 
A call back function is set to report the status to control module in case the download time exceeds the target download finish time. 
The control module schedules the video buffering when the call back function for target download time is triggered, 
the chunk download finishes, or the user swipes. 
Similar to Pensieve~\cite{Pensieve}, 
we also use an ABR server to run \systemname's algorithm on the same machine as the client. The control module communicates with the ABR server using XMLHttpRequests locally.

%% file: Eval.tex
\section{Evaluation}
\label{s:eval}

We evaluate \systemname{} across a variety of mobile 
network conditions, real user swipe traces, and videos. 
Our key findings:
\begin{itemize}
\item \systemname{} achieves QoE values within 77.3-98.6\% of the optimal, outperforming TikTok by 43.9-45.1$\times$.
\item In comparison to Tiktok, \systemname{} increases QoE rewards for video bitrates by 85.3-246.2\%, while also incurring 102.4-128.4$\times$ lower rebuffering penalties. Further, data wastage is 30\% lower with \systemname{} than TikTok.
\item \systemname{} is largely tolerant of errors in swipe distributions: with errors of 50\%, \systemname{} makes the correct buffering decisions 96.5\% of the time, yielding QoE reductions of only 10\% compared to when no distribution errors exist.

\end{itemize}

\subsection{Methodology}
\label{ss:env}

\begin{figure*}[htb]
\centering
  \begin{subfigure}[t]{0.33\linewidth}
      \centering
      \includegraphics[width=\linewidth]{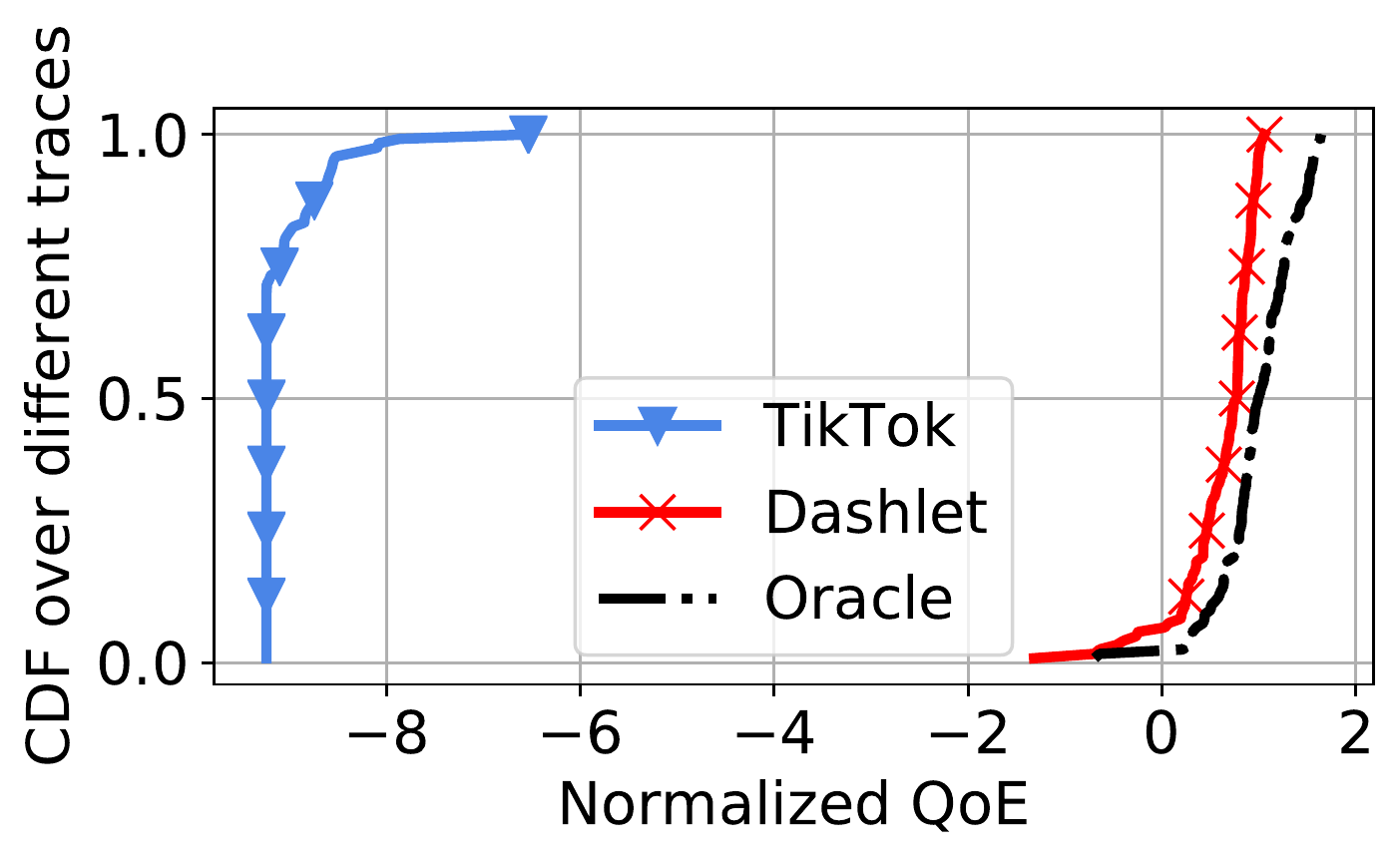} 
      \caption{Throughput < 1 Mbps.}
      \label{fig:qoe-fcc-low}
  \end{subfigure}\hfill
    \begin{subfigure}[t]{0.33\linewidth}
      \centering
      \includegraphics[width=\linewidth]{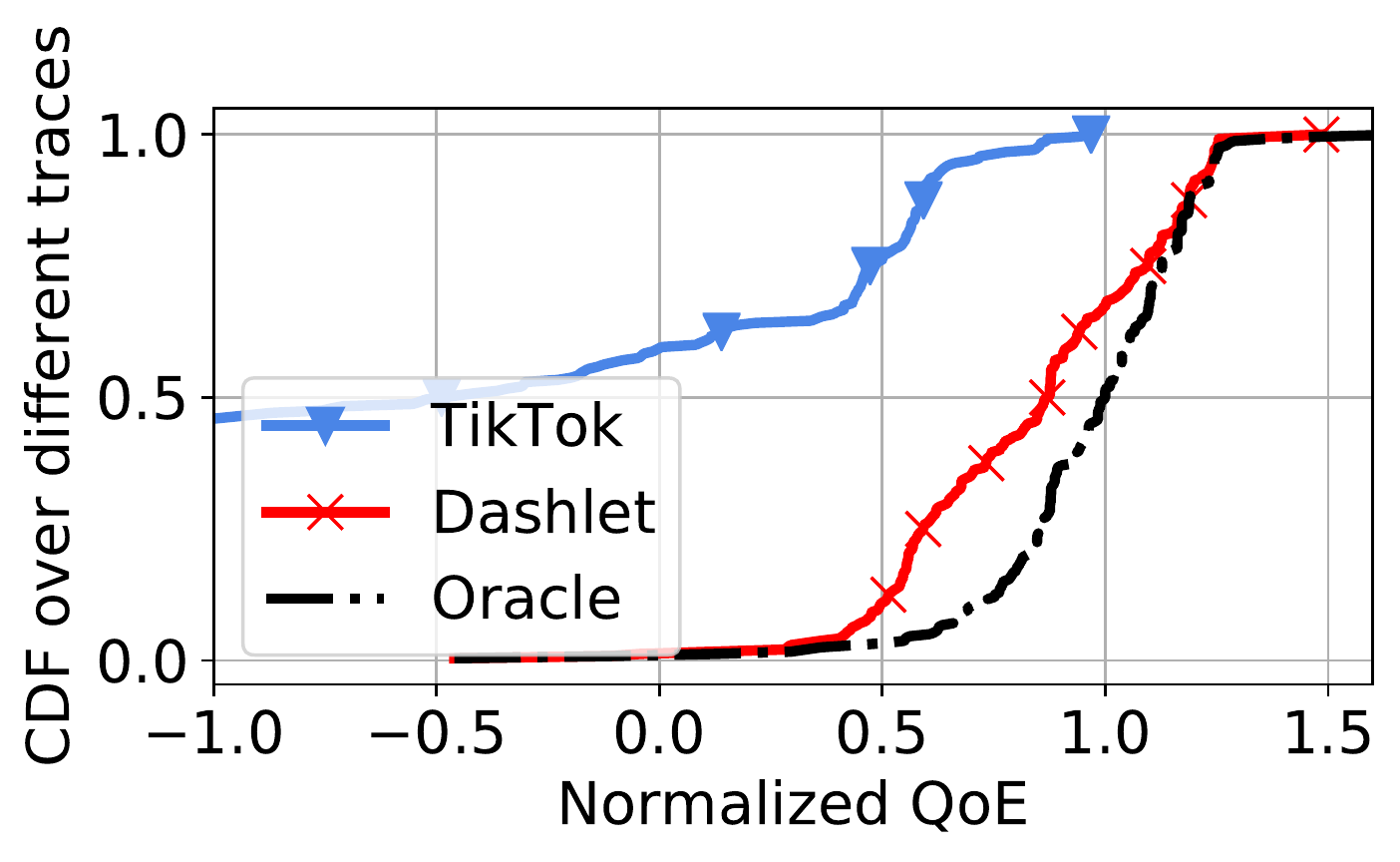} 
      \caption{1 Mbps < Throughput < 6 Mbps.}
      \label{fig:qoe-fcc-mid}
  \end{subfigure}\hfill
    \begin{subfigure}[t]{0.33\linewidth}
      \centering
      \includegraphics[width=\linewidth]{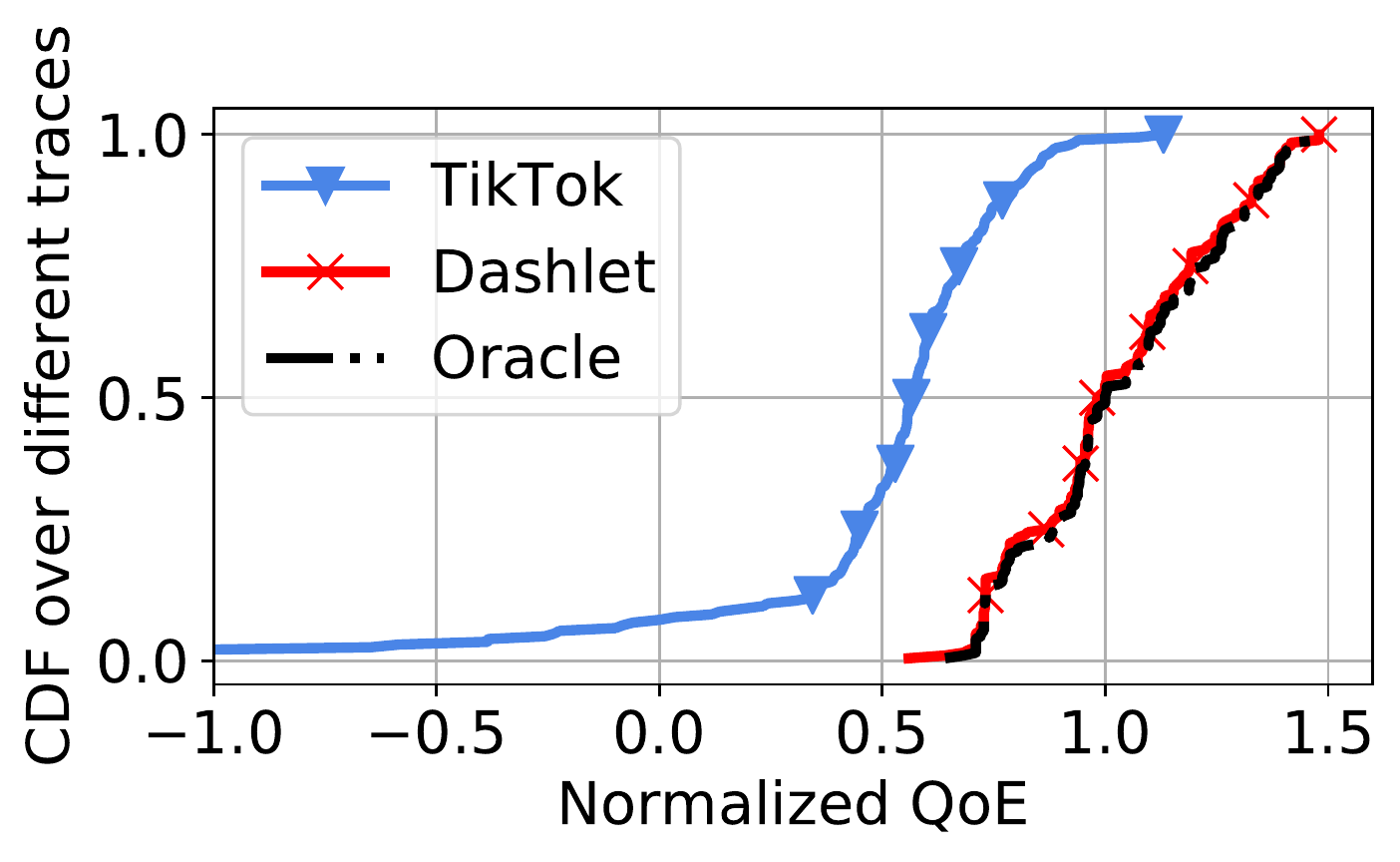} 
      \caption{Throughput > 6 Mbps.}
      \label{fig:qoe-fcc-high}
  \end{subfigure}\hfill
   \caption{\systemname\ \emph{vs.} TikTok and an Oracle scheme on the FCC LTE dataset and our entire corpus of swipe traces. Results are normalized to the median QoE of the Oracle algorithm in each case.}
   \label{fig:qoe-fcc}
    \vspace{-10pt}
\end{figure*}
\begin{figure*}[htb]
    \begin{minipage}[h]{0.33\linewidth}
      \centering
      \includegraphics[width=\linewidth]{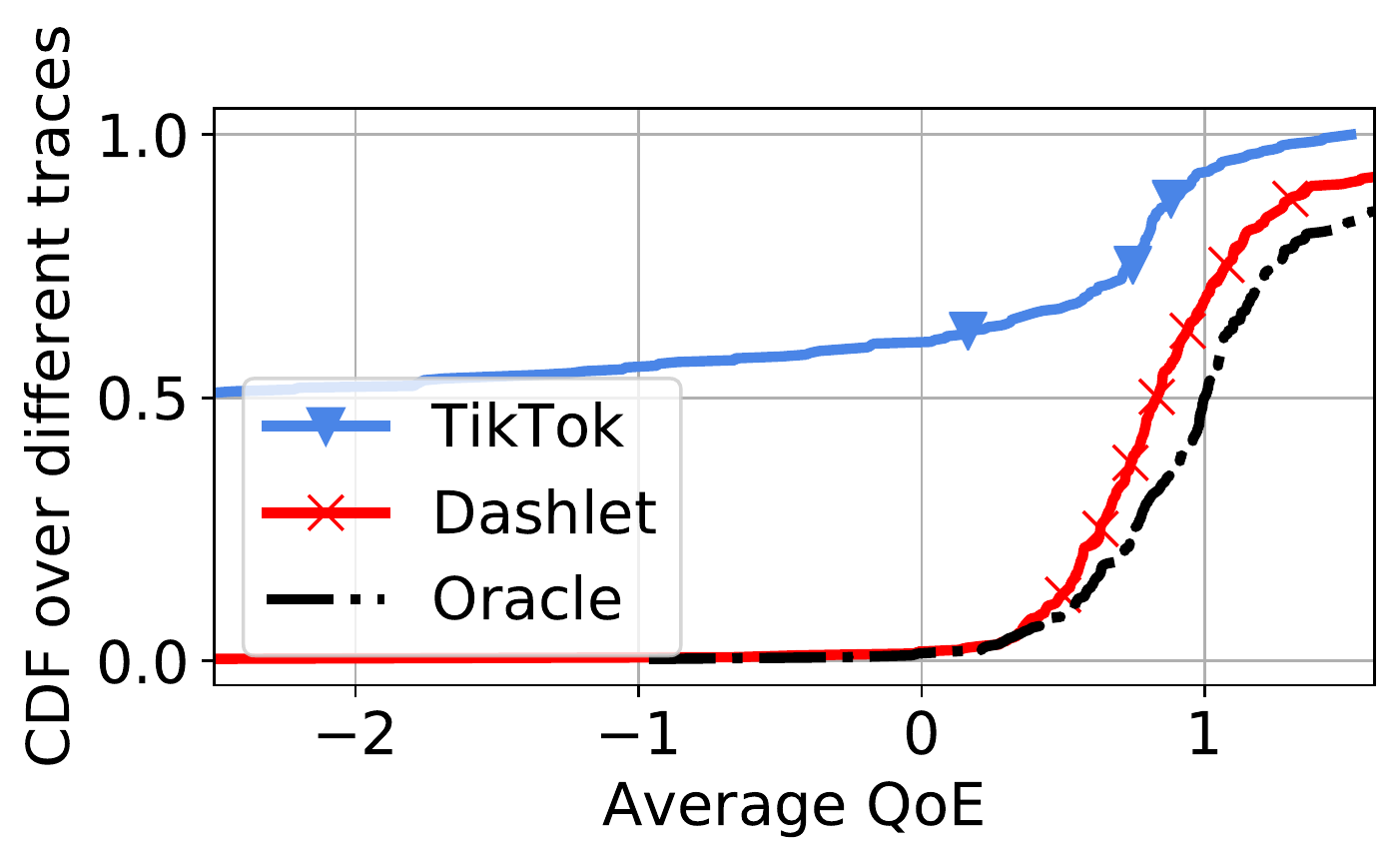} 
      \caption{ \systemname\ vs. TikTok and Oracle on a Mall WiFi dataset and our swipe traces.}
      \label{fig:qoe-wifi}
    \end{minipage}
    \hfill
    \begin{minipage}[h]{0.66\linewidth}
        \begin{subfigure}[t]{0.49\linewidth}
        \centering
      \includegraphics[width=\linewidth]{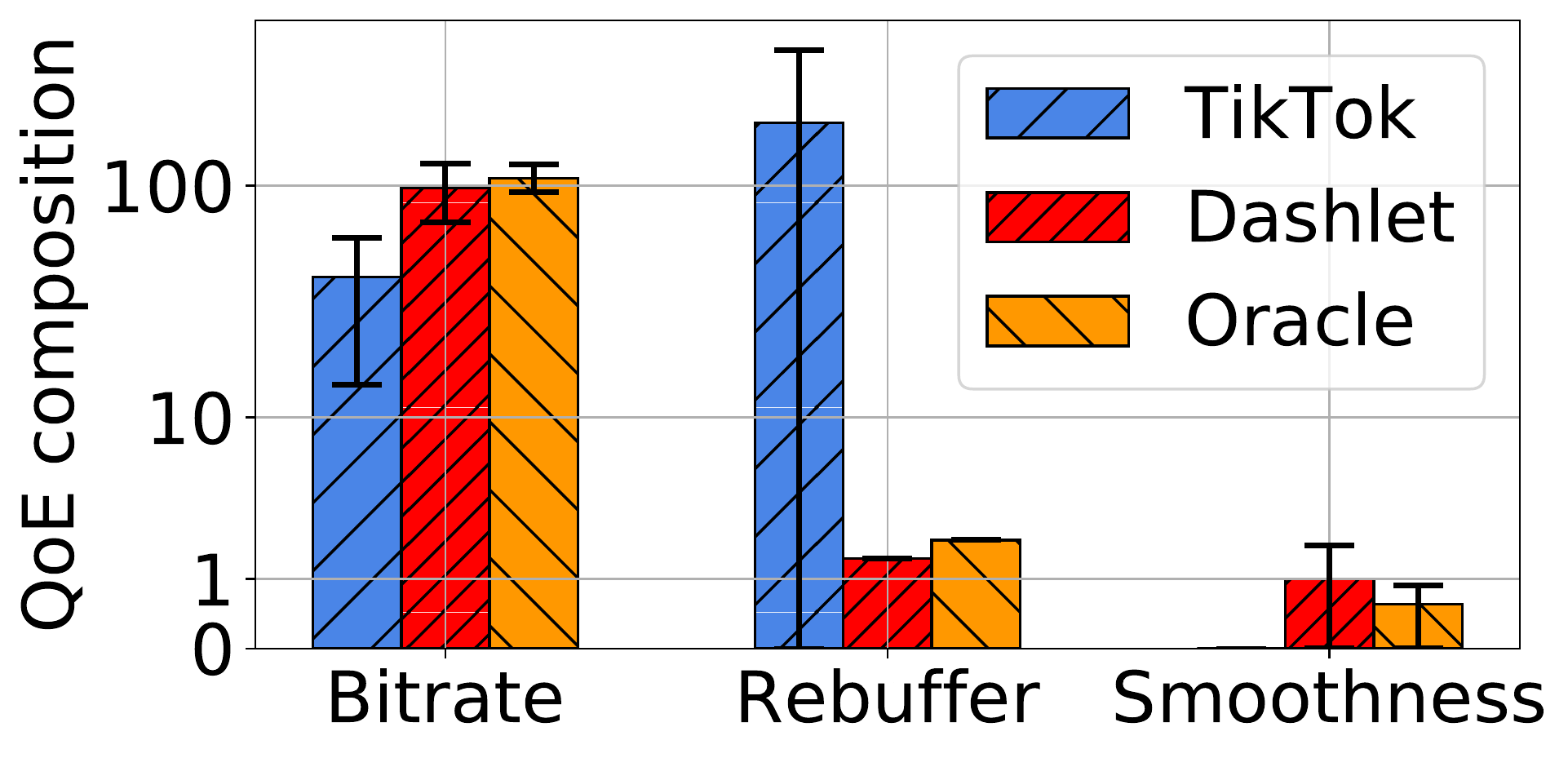} 
      \caption{FCC dataset (1-6~Mbps throughput).}
      \label{fig:decomposite-fcc}
  \end{subfigure}\hfill
    \begin{subfigure}[t]{0.49\linewidth}
      \centering
      \includegraphics[width=\linewidth]{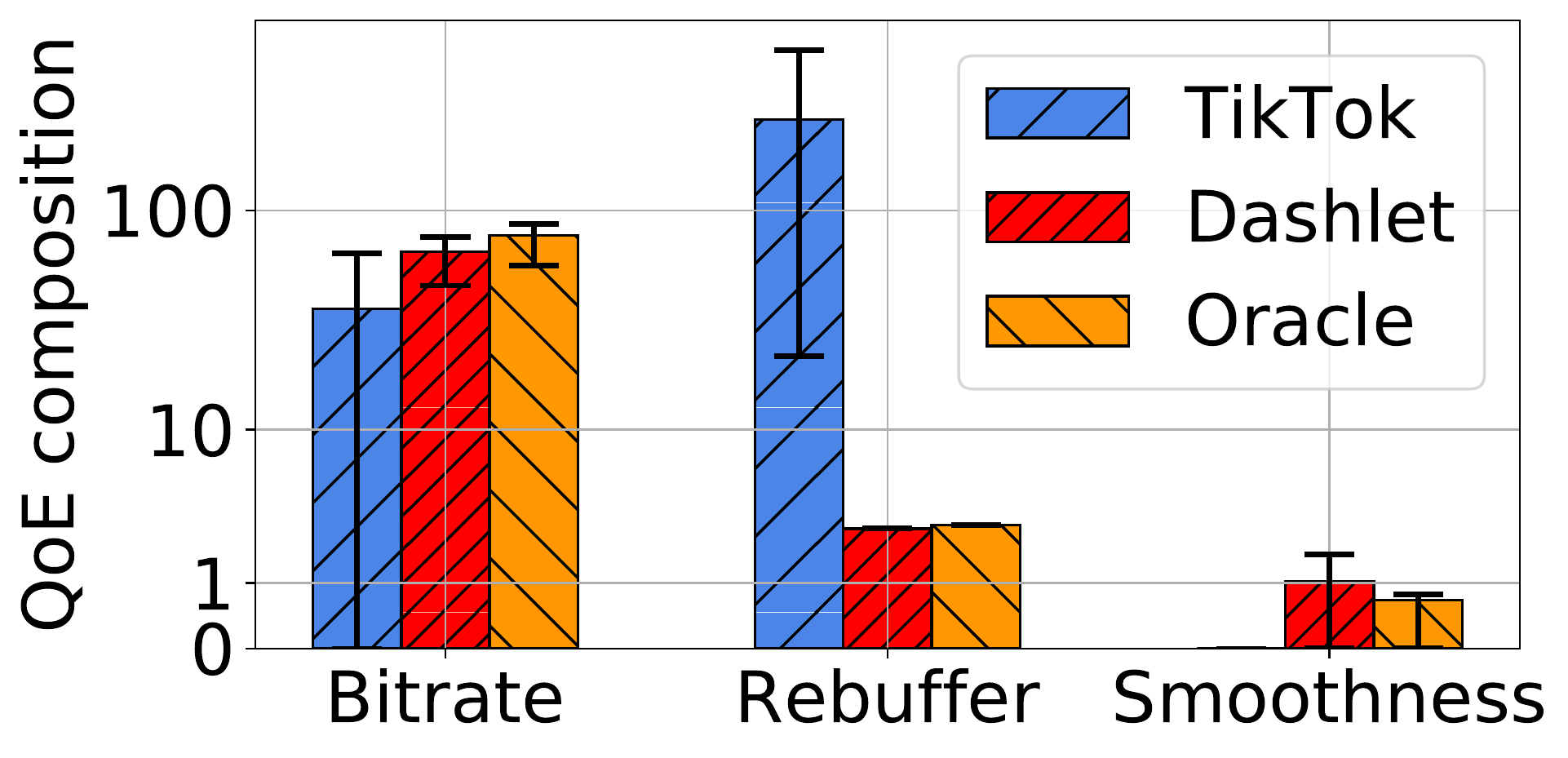} 
      \caption{Mall WiFi dataset.}
      \label{fig:decomposite-wifi}
  \end{subfigure}\hfill
   \caption{Breaking down QoE values in terms of bitrate reward, rebuffering penalty, and smoothness penalty. Bars list averages, with error bars for 1 st. dev. in each direction.}
   \label{fig:qoe-decomposititon}
    \end{minipage}
    \vspace{-10pt}
\end{figure*}

\paragraph{Baselines.}
We compare \systemname\ with the following systems:
\begin{itemize} 

\item \emph{TikTok:} we directly run the TikTok Android App (v. 15.7.46). To report QoE values and to analyze TikTok's behavior in our experiments, we employ the man-in-the-middle setup described in \S\ref{s:tiktok_analysis}.

\item \emph{Oracle:} we also run an `oracle' baseline that serves as an upper bound for QoE. The oracle is the RobustMPC algorithm~\cite{MPC} running with perfect (a priori) knowledge of both the user swipe traces and network throughput in each experiment. With that information, the algorithm knows the upcoming video viewing sequence at all times, and can thus pick the buffer sequences (and bitrates) that directly optimize QoE for the current network conditions.
\end{itemize}


\paragraph{Overall setup.} All video clients run on a rooted Pixel 2 phone (Android 10). The Oracle algorithm and \systemname{} run in the Google Chrome browser (v. 97.0.4692.87), and contact a local desktop which houses the videos accessed in each experiment (described below). In contrast, TikTok runs as an unmodified, native Android app and contacts Akamai CDNs to fetch video content as it normally does. All traffic to and from the phone passes over emulated mobile networks (which run atop WiFi connections with average speeds of $\approx$300 Mbps); to compensate for the discrepancy in video servers, we added 6 ms of round trip delay to traffic for \systemname{} and the Oracle algorithm, which reflects the maximum ping time we observed to the CDN used by TikTok.

\paragraph{Videos.} We consider 100 videos directly downloaded from TikTok's CDN server via keyword search for popular categories, e.g., racecar. Once downloaded, we use ffmpeg~\cite{ffmpeg} to transcode the videos into the DASH format for the same set of bitrates that TikTok supports (\S\ref{s:tiktok_primer}). By default, we store videos in 5-second chunks (which matches the average duration of the first chunks with TikTok when using the highest available bitrate); \S\ref{ss:chunksize} includes results for other chunk durations. Each experiment considers 10 minutes of viewing time to match the average session time for TikTok users~\cite{tiktokuser}. In order to enforce the same playing sequence (i.e., ordered list) of videos across the considered systems, we exploit the fact that the order in which videos are streamed with TikTok for a given keyword search remains unchanged on the order of many days. We use that same order across all systems and across experiments with different network and swipe traces.


\paragraph{Swipe traces.} We pair each video in our experiments with one used in our user studies (\S\ref{s:measurement-swipe}), trying to match video durations as closely as possible. When a video is played during an experiment, we replay a randomly-selected real user trace (i.e., exact swipe time) for the corresponding video in the user studies. Note that \systemname{} has access to the aggregated distribution of cross-user swipe times for each video, but does not have knowledge of the specific user trace (and thus, exact swipe times) that will be used in any experiment.


\paragraph{Network conditions.} We consider two sets of mobile network traces: (1) the FCC LTE dataset~\cite{fcc-trace} that is widely used in prior work~\cite{Pensieve,MPC}, and (2) a WiFi trace dataset that we collected in January 2022 in a shopping mall.


\paragraph{Evaluation metrics.} 
Short video streaming follows the same general goals as traditional video streaming scenarios~\cite{Pensieve,MPC}: maximize video bitrate, minimize rebuffering delays, and avoid frequent bitrate fluctuations. Thus, we adopt the following widely used QoE definition:
\begin{equation}
    QoE = R_{bitrate} - \mu \cdot P_{rebuffer} - \eta \cdot P_{smooth}
    \label{eq:qoe-define}
\end{equation}
where $R_{bitrate}$ is the average video bitrate, $P_{rebuffer}$ is the cumulative penalty for rebuffering (i.e., stalled playback), and $P_{smooth}$ is the penalty for frequent bitrate switching across adjacent chunks within a short video. We use the same values for $\mu$ and $\eta$ as prior work~\cite{Pensieve}, i.e., $\mu = 4.3$ and $\eta=1$. As discussed in \S\ref{s:intro}, user swipes can produce scenarios in which downloaded video is never watched by users. We compute QoE only based on viewed video chunks, but also report on the resources consumed for wasted downloads.



\subsection{End-to-End performance}
\label{ss:qoe-result}

Figure~\ref{fig:qoe-fcc} compares the average QoE of all three systems when running over the FCC LTE dataset. To aid analysis, results are broken down into 3 categories based on the available bandwidth in each trace. Figure~\ref{fig:qoe-wifi} shows the same results for the Mall WiFi traces. Reported QoE results are normalized to the median values for the Oracle in each case.


\begin{figure*}[htb]
    \centering
    \includegraphics[width=1.\linewidth]{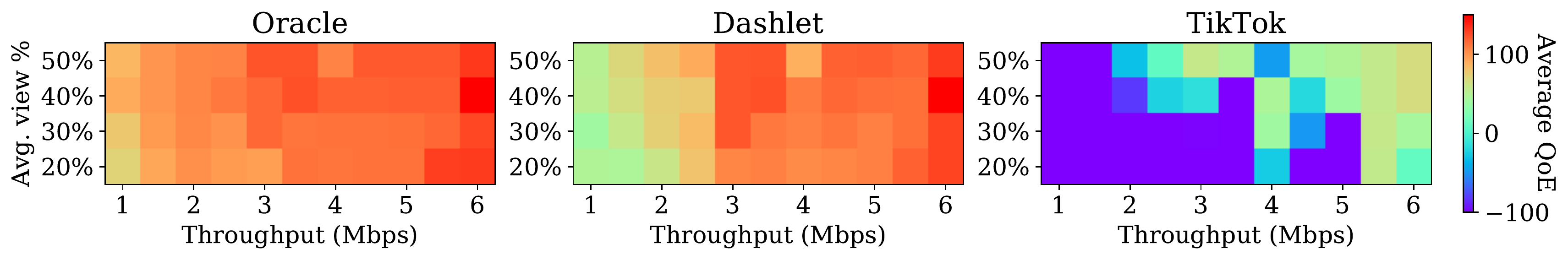}
    \caption{Average QoE with \systemname{}; results organized by average viewing percentage (based on swipe patterns) and network throughput.}
    \label{fig:swipe-network}
    \vspace{-10pt}
\end{figure*}

There are two key takeaways from these results. First, \systemname{} consistently outperforms TikTok across all settings, with QoE benefits that are 43.9-45.1$\times$ closer to the Optimal. For example, for the FCC LTE traces, (median) QoE values with \systemname{} are within 77.3-98.6\% of the median optimal values. TikTok's QoE values, on the other hand, has gaps of -923.6-56.7\% from the optimal across those cases. This performance advantage persists on WiFi, with percent differences from the optimal of 83\% and 266\% for \systemname{} and TikTok, respectively. The main reason is that, whereas \systemname{} buffers video according to swipe distributions and application playback constraints, TikTok follows a fixed algorithm in all cases that buffers many chunks that are never watched. The wasted resources translate to substantial rebuffering delays for required chunks (as discussed below), particularly in low-bandwidth scenarios, e.g., Figure~\ref{fig:qoe-fcc}(a). In addition, \systemname{}'s algorithm leaves the network idle only if the client-side playback buffer is fully occupied (and cannot house additional chunks); in contrast, TikTok enters a prebuffer-idling state after it completes prebuffering the first chunk for each video in the current group-of-ten (\S\ref{ss:tt-algo}).

Second, \systemname{}'s gap from the optimal shrinks as network throughput increases. For instance, on the FCC dataset, median gaps are 22.7\%, 12.7\%, and 1.4\% as the network ranges rise from $<$1 to 1-6 to $>$6 Mbps. The reason is that \systemname{} is able to grow and maintain larger playback buffers, which in turn lowers the risk of rebuffering from unanticipated swipes. In other words, the larger bandwidths mask the (already minimal) errors induced from errors in swipe distributions in terms of how well they characterize the current user.



\subsubsection{Breaking down QoE} 
\label{ss:qoe-decomposition}

Figure~\ref{fig:qoe-decomposititon} lists the contributing elements to overall QoE values of each system, i.e., bitrate reward, rebuffering penalty, and smoothness penalty. There are four points to note.

First, TikTok's performance degradations are primarily due to substantial rebuffering: its average rebuffering penalties in the QoE function are 186.5 and 259.2 for the LTE and WiFi datasets. This translates to upwards of 208 and 289 seconds of rebuffering in 10-minute video sessions, which is 128.4$\times$ and 102.4$\times$ higher than with \systemname{}. The root cause is \systemname{}'s conservative approach to explicitly minimize rebuffering in an effort to hedge against uncertainty in user swipe times. Second, and for the same reason, \systemname{}'s rebuffering penalty is lower than that with the Oracle algorithm. However, in doing so, \systemname{} also achieves moderately lower bitrate rewards compared to the Oracle. Third, compared to TikTok, \systemname{} achieves 246.2\% and 85.3\% higher video bitrates than TikTok on the FCC LTE and Mall WiFi datasets, respectively. Fourth, all three systems incur minimal smoothness penalties. This is because short videos last only 10s of seconds (and thus, a handful of chunks)). The algorithms can switch bitrates between videos to adapt to throughput changes without incurring smoothness penalties.

\subsection{Micro Benchmarks}

\subsubsection{Impact of Swipe and Network Speeds on QoE}
\label{ss:swipe-network}

Patterns in network throughput and user swipes largely influence the performance of short video streaming algorithms. To understand the effect of each, we report \systemname{}'s QoE results for different network throughputs (for our FCC LTE dataset) and swipe rates. As shown in Figure~\ref{fig:swipe-network}, the major factor that affects QoE with \systemname\ is the network throughput. Importantly, swipe speed does not have a significant impact on \systemname{}'s performance, validating its use of only coarse information from swipe distributions to hedge against different swipe patterns. In contrast, both network throughput and swipe speed have a large impact on TikTok's QoE.

\subsubsection{Network Idle and Data Waste} 
\label{ss:idle-waste}

\begin{figure*}[t]
    \begin{minipage}[h]{0.29\linewidth}
    \centering
    \includegraphics[width=0.85\linewidth]{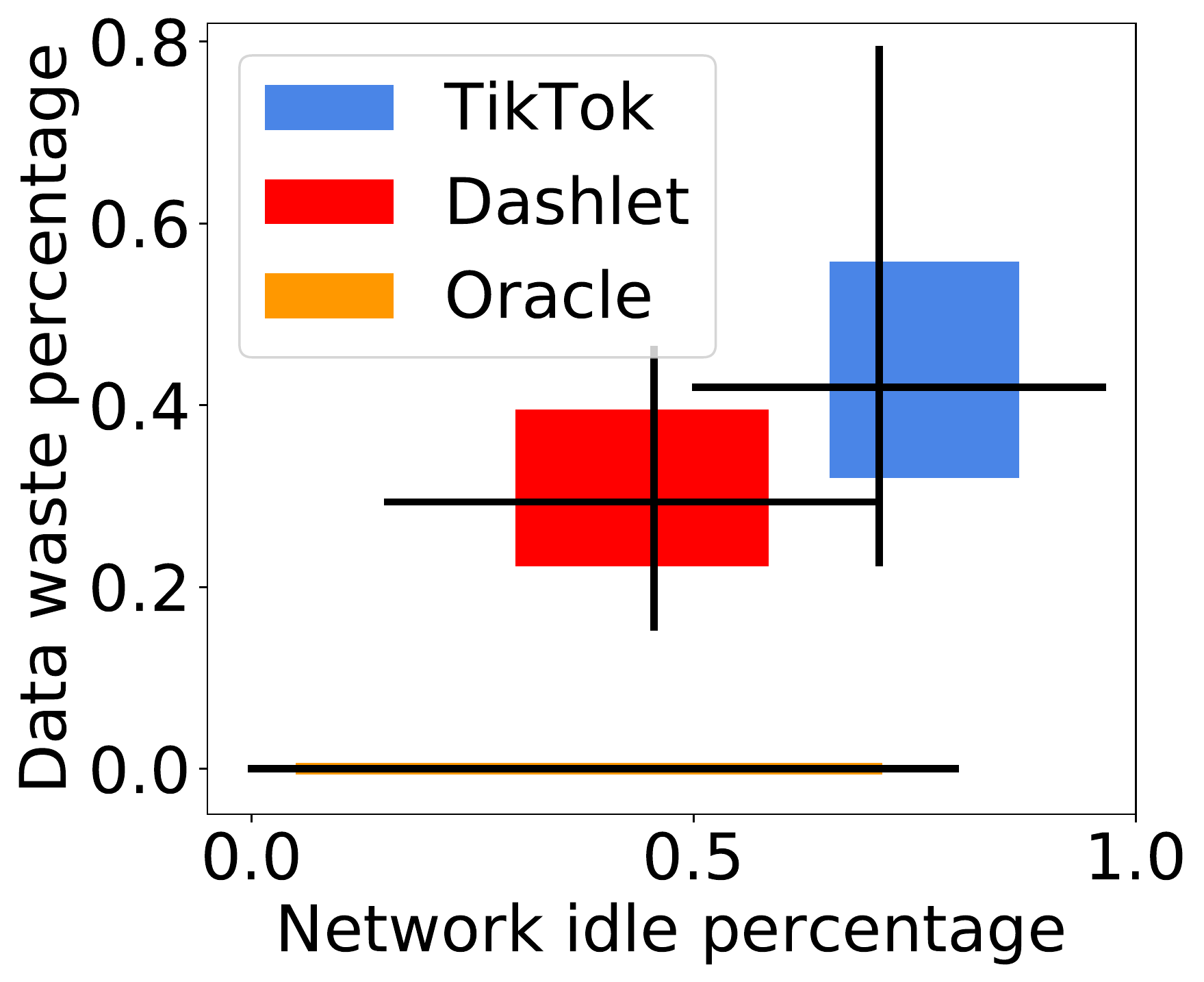}
    \caption{Data wastage and network idle time. Boxes span 25-75th percentiles. Black lines span min/max, and intersect at the median for both properties.}
    \label{fig:waste-idle}
    \end{minipage}
    \hfill
    \begin{minipage}[h]{0.33\linewidth}
    \centering
        \includegraphics[width=1.0\linewidth]{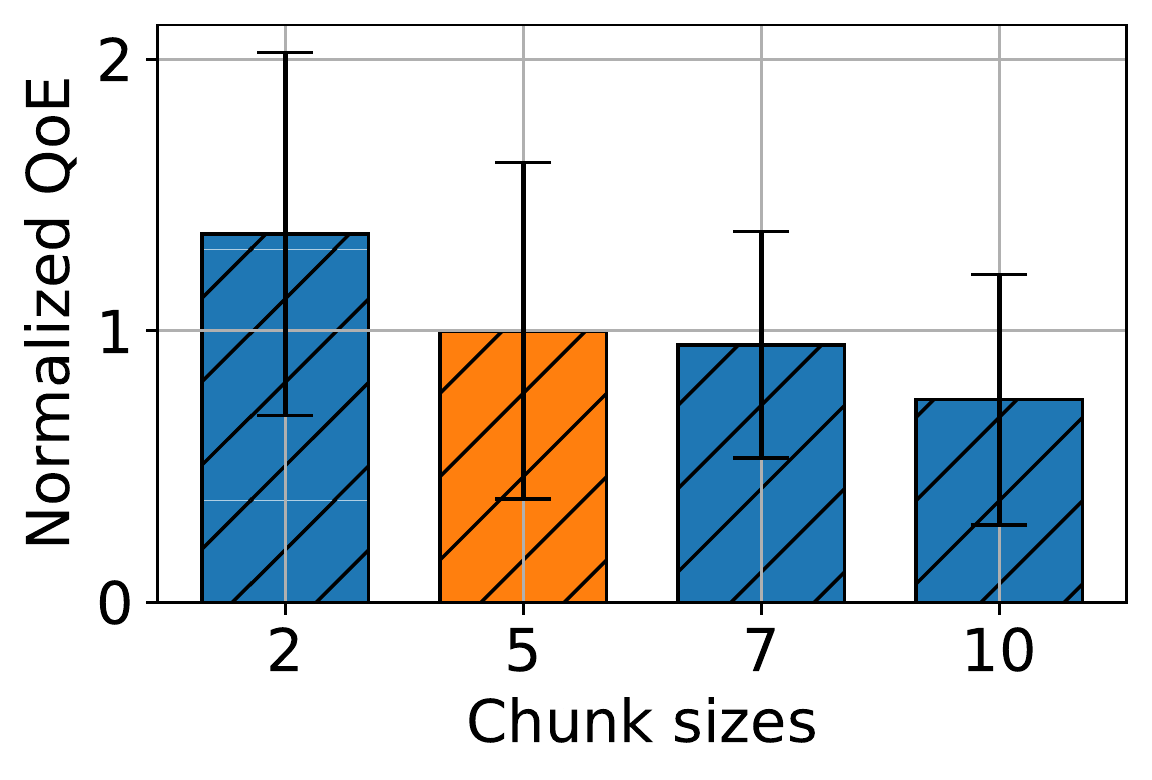}
    \caption{Chunk size's impact on QoE. Bars list averages, with error bars for one st. dev. in each direction.}
    \label{fig:qoe-chunks}
    \end{minipage}
    \hfill
    \begin{minipage}[h]{0.33\linewidth}
    \centering
    \includegraphics[width=1.0\linewidth]{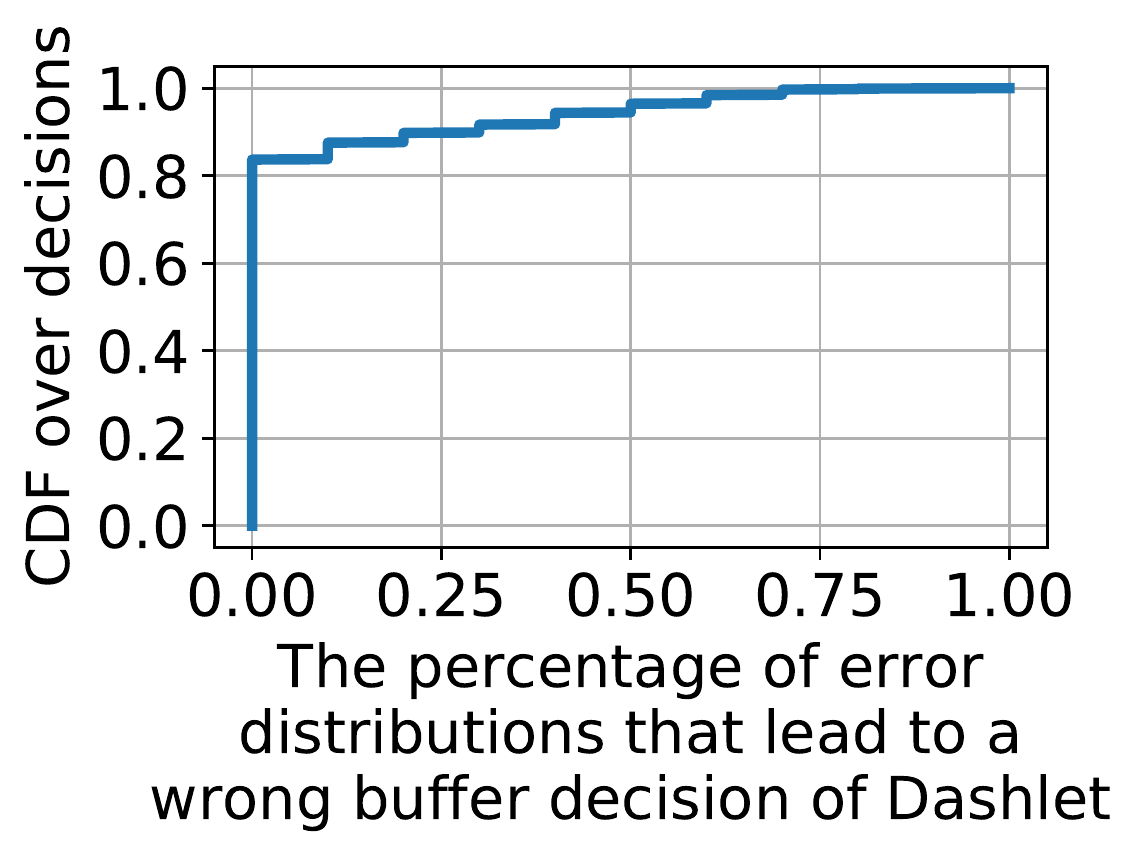}
    \caption{\systemname{}'s tolerance to swipe distribution errors.}
    \label{fig:swipe-stable}
    \end{minipage}
    \vspace{-10pt}
\end{figure*}

To dig deeper into \systemname{}'s higher QoE values over TikTok from \S\ref{ss:qoe-decomposition}, we investigate network idling and data wastage for both systems. 
Figure~\ref{fig:waste-idle} shows our results; note that the Oracle algorithm does not incur any data wastage since it has perfect knowledge of user swipe times. As shown, median data wastage and idle time for \systemname\ are 29.4\% and 45.5\%, respectively, which are 30.0\% and 35.9\% lower than those with TikTok. These improvements, in turn, enable \systemname{} to stream video at higher bitrates than TikTok for a given network throughput while keeping rebuffering delays low.

\subsubsection{The impact of chunk size on \systemname's QoE} 
\label{ss:chunksize}

Unlike TikTok which breaks up videos by bytes, \systemname{} (by default) breaks up videos into 5-second chunks. We evaluated the impact that chunk sizes have on \systemname{}'s performance by considering the following chunk sizes (based on prior work~\cite{zhang2017modeling}): \{2, 5, 7, 10\} seconds. Note that we did not modify chunk sizes for TikTok as we could not alter its video servers. As shown in Figure~\ref{fig:qoe-chunks}, \systemname{}'s performance steadily decreases as chunk sizes grow, \eg, average QoE drops by 35.4\% as chunk sizes grow from 5 to 10 seconds. The reason is that data wastage grows with larger chunk sizes: a user swipe at 1 second into a chunk will result in more wasted bytes with chunks of 10 seconds than 2 seconds.


\subsubsection{Decision Stability with Swipe Prediction Errors}
\label{ss:decision-stable}

\systemname\ determines buffer sequences by leveraging (coarse information from) users' swipe distributions for each video. Thus, a natural question is how robust are \systemname{}'s decisions to errors in those distributions, i.e., does it make the right decisions even with different degrees of errors?


Recall that there are three inputs to \systemname{}'s algorithm at any time: the swipe distribution for each considered video, the estimated network throughput, and the client-side player's current buffer state. The algorithm uses this information and returns a buffer sequence of chunks to download, with the first chunk in the ordered list indicating the action to perform immediately, i.e., the chunk to download now. To answer the above question, we profiled the above inputs throughout our experiments, and then compared the actions selected by \systemname\ with those that it would select if the input swipe distribution involved errors. In particular, we considered 10 versions of each video's distribution by (roughly) modeling its original distribution as an exponential one, and then altering the corresponding $\lambda$ value to change the average swipe time by 1$\pm$\{0-50\%\} (in 10\% increments).

Figure~\ref{fig:swipe-stable} shows our results. As shown, 83.7\% of \systemname{}'s decisions are unchanged across all of the considered distribution errors. The values remain relatively stable as errors grow -- \eg, 96.5\% of \systemname{}'s decisions are unchanged with errors of 50\% -- but begin to grow after 82\%. These results illustrate that \systemname{} only relies on coarse information from swipe distributions (\eg, about whether a user is likely to swipe early or late in the video); it is for this reason that decisions are varied only when errors are very high (and even the coarse information that \systemname{} uses has changed).






\begin{figure}[htb]
    \begin{minipage}[h]{0.49\linewidth}
    \centering
    \includegraphics[width=1.0\linewidth]{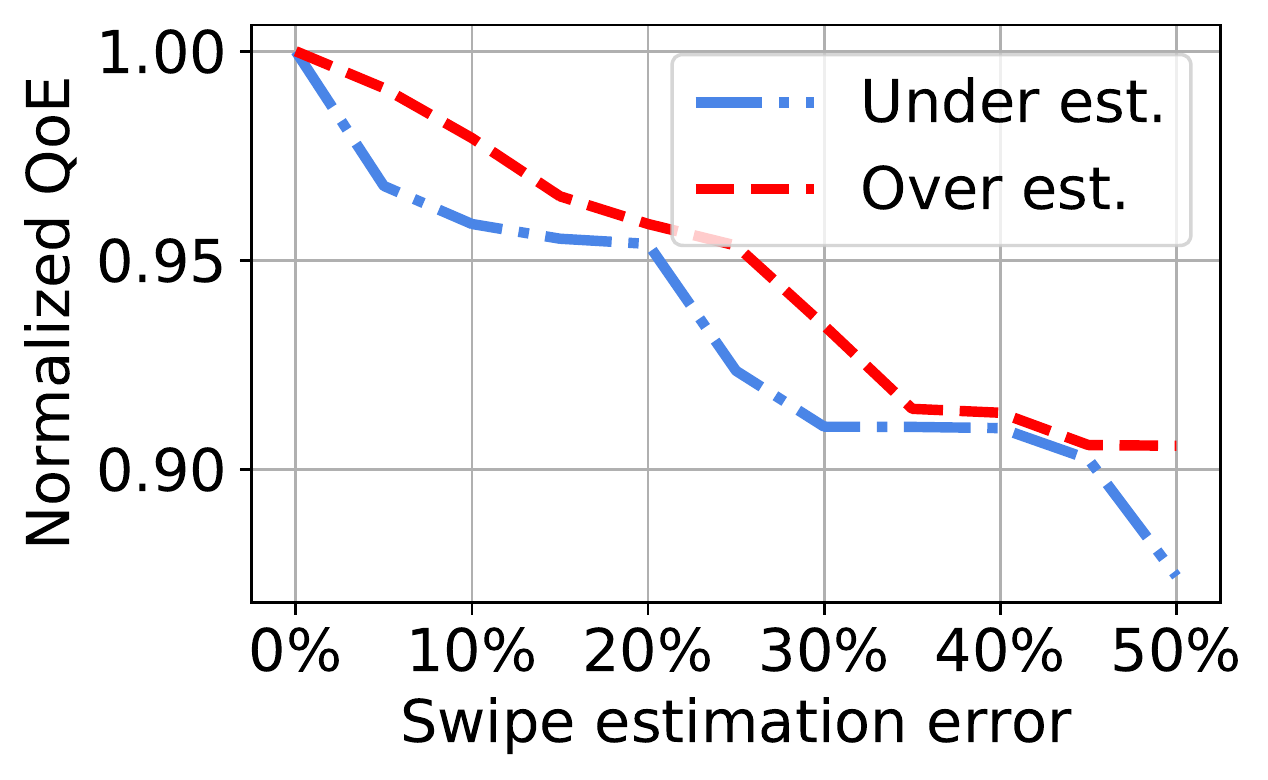}
    \caption{Impact of swipe estimation errors on \systemname.}
    \label{fig:impact-swipe-estimation}
    \end{minipage}
    \hfill
    \begin{minipage}[h]{0.49\linewidth}
    \centering
    \includegraphics[width=1.0\linewidth]{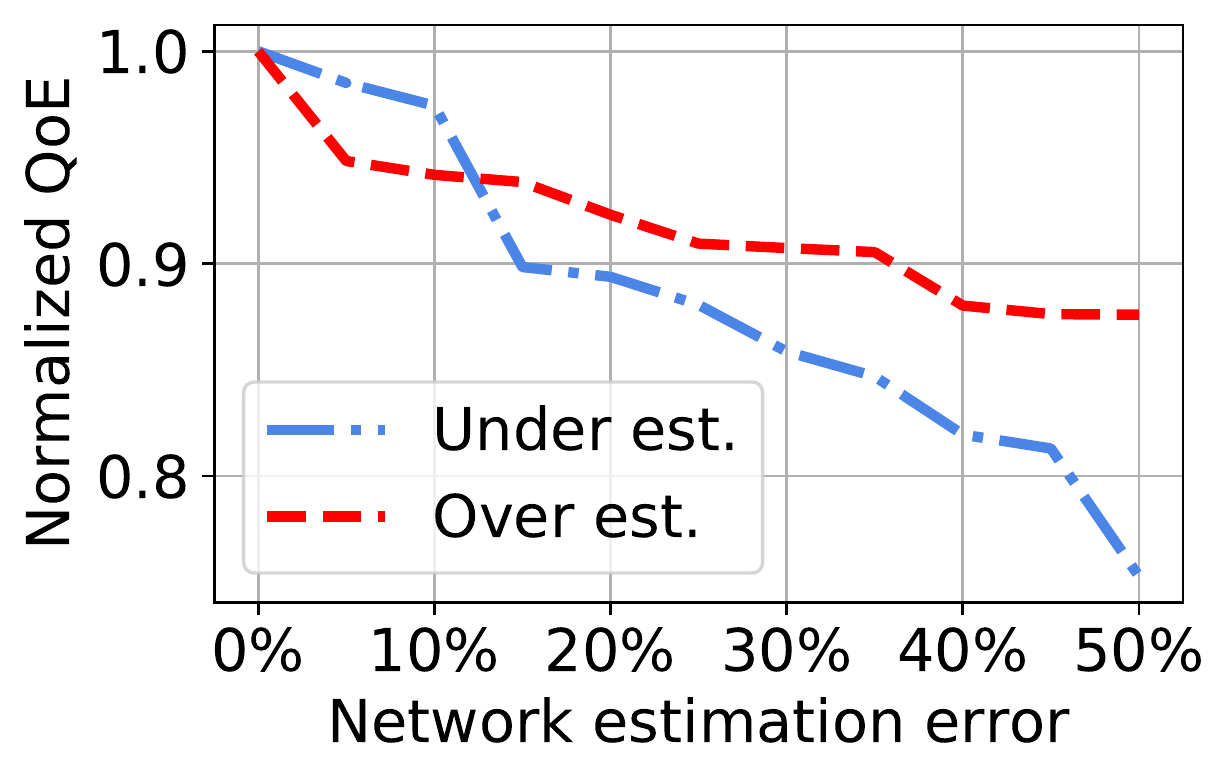}
    \caption{Impact of network estimation errors on \systemname.}
    \label{fig:impact-net-estimation}
    \end{minipage}
    \vspace{-10pt}
\end{figure}

\subsubsection{QoE sensitivity with Swipe and Network Errors} 
\label{ss:swipe-error}

Building on the previous results, we now analyze how errors in swipe distribution affect the QoE that \systemname{} delivers. Using same faulty distributions from above, we ran \systemname{} on all videos and the network traces in our FCC LTE dataset. Figure~\ref{fig:impact-swipe-estimation} shows the results, breaking them down in terms of scenarios with over estimation of swipe times (longer average viewing time than the correct distribution, i.e., later swipes) and under estimation (shorter average viewing time). As shown, \systemname{} is quite tolerant to such errors, delivering 87\% and 91\% of its full QoE (with no errors) when the traces are over- and under-estimating swipe times by 50\%.

We perform a similar analysis to evaluate \systemname{} in the presence of network prediction errors. Specifically, we replace the network predictor in RobustMPC~\cite{MPC} with one that reads in the actual instantaneous throughput from the current Mahimahi trace, and multiplies that value by between 1$\pm$\{0-50\%\}. Overall, as per Figure~\ref{fig:impact-net-estimation}, we find that \systemname{}'s QoE drops to 88\% and 76\% of its values without network errors when the network estimate is over- or under-estimating by 50\%. These results highlight that \systemname\ is more robust to errors in swipe distributions than network forecasts.

%% file: related.tex
\section{Related work}
\label{s:background}

\paragraph{Traditional adaptive video streaming} Traditional video streaming services deliver video content from the CDN to the user with adaptive bitrate system with the objective of maximizing the quality of experience for users~\cite{krishnan2013video}. Research effort has been made to improve the quality of experience from different perspectives, including 
streaming algorithm~\cite{MPC,spiteri2020bola,Pensieve,jiang2012improving,kim2020neural},
video codec~\cite{fouladi2018salsify,dasariswift}, network prediction~\cite{puffer,sun2016cs2p}, protocol design~\cite{zheng2021xlink,han2016mp}, 
and video super resolution~\cite{yeo2018neural,yeo2020nemo}. But all these optimization is for the same video streaming model: the video download sequence is the same as the video playing sequence. \systemname\ also uses QoE as the optimization goal but tackles a different problem as in the short video streaming the video download sequence is the same as the video playing sequence due to users' swipes.

\paragraph{Streaming new form of video} There are also rising interest on 360 degree video~\cite{qian2018flare, guan2019pano} and volumetric video streaming~\cite{qian2019volumetric}. These systems need to handle the uncertainty from the users' head position or location. \systemname's design also models the uncertainty from the user swipe patterns. But \systemname\ targets on a different problem compared with 360 degree or volumetric video streaming. Some existing works~\cite{ran2020ssr,he2020liveclip} also try to apply reinforcement learning algorithms from traditional video streaming~\cite{Pensieve} to short video streaming. However, these works do not factor in the impact of user swipes on buffering decisions as \systemname{} does. 





%% file: concl.tex
\section{Conclusion}
\label{s:concl}
In this paper, we design and implement \systemname\ with the insight provided by measurement for a commercial short video app and a user  study on general user swipe pattern. \systemname's algorithm strategically determines the buffer order with the input from a coarse-grained swipe distribution. Evaluation result shows  \systemname\ significantly improves  video quality and reduces  rebuffering compared with the baseline system.
